\RequirePackage{fix-cm}
\documentclass[twocolumn,epjc3]{svjour3}  
\RequirePackage[T1]{fontenc}
\smartqed  
\usepackage[utf8]{inputenc}
\RequirePackage{graphicx}
\RequirePackage{mathptmx}      
\RequirePackage{flushend}
\RequirePackage[colorlinks,citecolor=blue,urlcolor=blue,linkcolor=blue]{hyperref}
\usepackage[separate-uncertainty,retain-explicit-plus,per-mode=symbol,binary-units]{siunitx}
\usepackage[version=4]{mhchem}

\usepackage{enumitem}
\setdescription{itemsep=0pt,parsep=0pt,labelsep=.2cm,leftmargin=2.2cm,labelwidth=2cm,listparindent=.7cm}
\setlist{nolistsep,leftmargin=1cm}
\newlist{enumcompactitem}{itemize}{3}
\setlist[enumcompactitem]{topsep=0pt,partopsep=0pt,itemsep=0pt,parsep=0pt}
\setlist[enumcompactitem,1]{label=\textbullet}
\setlist[enumcompactitem,2]{label=---}
\setlist[enumcompactitem,3]{label=*}
\newlist{enumcompactdesc}{description}{3}
\setlist[enumcompactdesc]{topsep=0pt,partopsep=0pt,itemsep=0pt,parsep=0pt}
\newlist{enumcompactenum}{enumerate}{3}
\setlist[enumcompactenum]{topsep=0pt,partopsep=0pt,itemsep=0pt,parsep=0pt}
\setlist[enumcompactenum,1]{label=\arabic*}
\setlist[enumcompactenum,2]{label=\alph*}
\setlist[enumcompactenum,3]{label=\roman*}
\DeclareSIUnit\c{\mbox{$c$}}
\DeclareSIUnit\magn{\mbox{$\times$}}
\DeclareSIUnit\min{min}
\DeclareSIUnit\week{week}
\DeclareSIUnit\month{mo}
\DeclareSIUnit\months{mos}
\DeclareSIUnit\year{yr}
\DeclareSIUnit\years{years}
\DeclareSIUnit\yr{yr}
\DeclareSIUnit\standard{std}
\DeclareSIUnit\str{sr}
\DeclareSIUnit\ppm{ppm}
\DeclareSIUnit\ppb{ppb}
\DeclareSIUnit\ppt{ppt}
\DeclareSIUnit\pe{PE}
\DeclareSIUnit\spe{SPE}
\DeclareSIUnit\pdm{PDM}
\DeclareSIUnit\ev{events}
\DeclareSIUnit\ct{counts}
\DeclareSIUnit\neutron{\mbox{$n$}}
\DeclareSIUnit\smp{samples}
\DeclareSIUnit\Sample{S}
\DeclareSIUnit\ch{ch}
\DeclareSIUnit\hit{hit}
\DeclareSIUnit\hits{hits}
\DeclareSIUnit\bin{(\mbox{5-PE}~bin)}
\DeclareSIUnit\sgm{\mbox{$\sigma$}}
\DeclareSIUnit\rms{RMS}
\DeclareSIUnit\keVee{\mbox{keV$_{e{\rm e}}$}}
\DeclareSIUnit\keVr{\mbox{keV$_{\rm nr}$}}
\DeclareSIUnit\eVee{\mbox{eV$_{\rm ee}$}}
\DeclareSIUnit\eVr{\mbox{eV$_{\rm nr}$}}
\DeclareSIUnit\ph{photon}
\DeclareSIUnit\el{\mbox{$e^-$}}
\DeclareSIUnit\pm{\mbox{PMT}}
\DeclareSIUnit\pixel{\mbox{pixel}}
\DeclareSIUnit\inch{''}
\DeclareSIUnit\foot{'}
\DeclareSIUnit\bit{bit}
\DeclareSIUnit\sample{samples}
\DeclareSIUnit\barn{barn}
\DeclareSIUnit\bara{bar}
\DeclareSIUnit\barg{barg}
\DeclareSIUnit\mlardepth{\mbox(meter~of~\LAr~depth)}
\DeclareSIUnit\Curie{Ci}
\DeclareSIUnit\psf{psf}
\DeclareSIUnit\pcf{pcf}
\DeclareSIUnit\parsec{pc}
\DeclareSIUnit\mwe{\mbox{m.w.e.}}
\DeclareSIUnit\liveday{\mbox{live-days}}
\DeclareSIUnit\days{\mbox{days}}
\DeclareSIUnit\miles{\mbox{miles}}
\DeclareSIUnit\lumens{\mbox{lm}}
\DeclareSIUnit\degreeC{\mbox{$^{\circ}$C}}
\DeclareSIUnit\degreeF{\mbox{$^{\circ}$F}}
\DeclareSIUnit\electron{\mbox{$e^-$}}
\DeclareSIUnit\Euro{\mbox{\euro}}
\DeclareSIUnit\cph{cph}
\DeclareSIUnit\neq{neq}
\DeclareSIUnit\normal{\mbox{N}}

\newcommand{\ReD}{\mbox{ReD}}

\newcommand{\TPC}{\mbox{TPC}}

\newcommand{\LArTPC}{\mbox{LAr~TPC}}

\newcommand{\SiPM}{\mbox{SiPM}}
\newcommand{\SiPMs}{\mbox{SiPMs}}

\newcommand{\LAr}{\ce{LAr}}

\newcommand{\NR}{\mbox{NR}}
\newcommand{\NRs}{\mbox{NRs}}

\usepackage{siunitx}
\usepackage{lineno}
\usepackage{textcomp}
\usepackage[dvipsnames]{xcolor}
\usepackage{booktabs}
\usepackage{makecell}
\usepackage{amsmath}
\usepackage{amssymb}
\renewcommand{\NR}{\mbox{NR}}

\renewcommand{\ReD}{\mbox{ReD}}
\renewcommand{\LArTPC}{\mbox{LAr~TPC}}
\renewcommand{\TPC}{\mbox{TPC}}

\renewcommand{\SiPM}{\mbox{SiPM}}
\renewcommand{\SiPMs}{\mbox{SiPMs}}
\newcommand{\Am}{$^{241}$Am}

\newcommand{\Li}{$^{7}$Li}
\newcommand{\Be}{$^{7}$Be}
\newcommand{\Er}{$E_r$}

\newcommand{\NexNi}{$N_\mathrm{ex}/N_\mathrm{i}$}

\newcommand{\fprompt}{$f_\mathrm{p}$}
\newcommand{\tdrift}{$t_\mathrm{drift}$}
\newcommand{\ISO}[2]{$^{#2}\mathrm{#1}$}
\newcommand{\edrift}{$\mathcal{E}_d$}
\newcommand{\eex}{$\mathcal{E}_{ex}$}
\newcommand{\eel}{$\mathcal{E}_{el}$}

\newcommand{\xy}{$x-y$}
\newcommand{\vect}[1]{\boldsymbol{#1}}

\journalname{Eur. Phys. J. C}
\makeatletter
\renewcommand{\thanksref}[1]{\nolinebreak\textsuperscript{\ref{#1}}\nolinebreak\checknextarg}
\newcommand{\checknextarg}{\@ifnextchar\bgroup{\nolinebreak\gobblenextarg}{}}
\newcommand{\gobblenextarg}[1]{ \textsuperscript{\nolinebreak\hspace{-4pt}\mbox{\nolinebreak$^,$
\nolinebreak\ref{#1}\nolinebreak}\nolinebreak} \@ifnextchar\bgroup{\gobblenextarg}{}}

\begin{document}

\title{Directionality of nuclear recoils in a liquid argon time projection chamber
}



\author{The DarkSide-20k Collaboration$^\text{\normalfont a,1}$}

\thankstext{e1}{e-mail: ds-ed@lngs.infn.it}

\institute{See back for author list \label{addr1}}



\date{Received: date / Accepted: date}

\maketitle

\begin{abstract}
The direct search for dark matter in the form of weakly interacting massive particles (WIMP) is performed by detecting nuclear recoils (NR) produced in a target material from the WIMP elastic scattering. 
A promising experimental strategy for direct dark matter search employs argon dual-phase time projection chambers (\TPC). One of the advantages of the \TPC\ is the capability to detect both the scintillation and charge signals produced by NRs. 
Furthermore, the existence of a drift electric field in the \TPC\ breaks the rotational symmetry: the angle between the drift field and the momentum of the recoiling nucleus can potentially affect the charge recombination probability in liquid argon and then the relative balance between the two signal channels.
This fact could make the detector sensitive to the directionality of the WIMP-induced signal, enabling unmistakable annual and daily modulation signatures for future searches aiming for discovery. The Recoil Directionality (\ReD) experiment was designed to probe for such directional 
sensitivity. The \TPC\ of \ReD\ was irradiated with neutrons at the INFN Laboratori Nazionali del Sud, and data were taken with \SI{72}{\kilo\electronvolt} NRs of known recoil directions. The direction-dependent liquid argon charge recombination model by Cataudella et al. was adopted and a likelihood statistical analysis was performed, which gave no indications of significant dependence of the detector response to the recoil direction. The aspect ratio $R$ of the initial ionization cloud is estimated to be $1.037\pm0.027$ and the upper limit is $R < 1.072$ with \SI{90}{\percent} confidence level. 
\end{abstract}

\section{Introduction}\label{intro}
A range of evidences from astronomy and cosmology~\cite{Ade:2016bk,Rubin:1999hb,Clowe:2006hr,Covone:2009ji,Malaney:1993ep} indicates that a substantial fraction of the Universe is made of non-baryonic dark matter, whose nature is still unknown. 
Weakly interacting massive particles (WIMPs), a common candidate, are actively searched for by many experiments worldwide using different technologies~\cite{Schumann:2019eaa,Cebrian:2022brv,Roszkowski:2017nbc}.  
The expected signal of those direct dark matter experiments is the nuclear recoil (NR) induced by the WIMP elastic scattering, having energy up to a few tens of \si{\kilo\electronvolt}. To improve sensitivity, it is crucial
to strongly suppress any NR generating contamination, until the ``neutrino fog'' is reached, i.e. the irreducible background from coherent elastic neutrino-nucleus scattering.

While evidence of WIMPs could be claimed based on an excess of NRs with respect to the expected background, a convincing discovery requires the observation of the effect in many target materials and the consistency with strong and unmistakable dark matter signatures. The motion of the Solar system relative to the galactic dark matter halo creates an apparent WIMP flux through terrestrial detectors coming from the direction opposite to the Earth's velocity vector, i.e. approximately from the Cygnus constellation. Furthermore, the motion of the Earth around the Sun generates annual modulations in the flux strength and direction. These signatures can be detected in the angular distribution of NRs produced by WIMP elastic scattering in a terrestrial detector, which would be an unmistakable ``smoking gun'' for dark matter, as none of the known backgrounds, including coherent neutrino scattering, can mimic it. 

Directional sensitivity would hence be a crucial asset for future direct dark matter search experiments, especially when claiming a signal. Even in a detector with moderate angular resolution ($\sim \SI{20}{\degree}$) for NRs, a few hundreds of events would be sufficient to reject the hypothesis of isotropic incident flux at $3 \sigma$ level~\cite{Cadeddu:2017ebu}. A number of R\&D programs is currently
in progress for directional direct dark matter search~\cite{Ahlen:2009ev,Hochberg:2016ntt,Battat:2016pap,Vahsen:2020pzb,Belli:2022jqq}.

One of the most promising current approaches for the direct search of WIMPs is based
on argon dual-phase Time Projection Chambers (\TPC).
It offers extremely low background thanks to the efficient rejection of
the electron recoil (ER) background provided 
by pulse shape discrimination (PSD)~\cite{Amaudruz:2016dq} and the use of low-radioactivity argon from underground 
sources~\cite{Galbiati:2007xz,Aalseth:2020nwt}. Based on the successful experience of the
DarkSide-50 experiment~\cite{Agnes:2015gu,Agnes:2015ftt,Agnes:2018ves} at the Gran Sasso Laboratory (LNGS) of
INFN and of the DEAP-3600 experiment~\cite{DEAP-3600:2017uua,DEAP:2019yzn} at SNOLAB, the Global Argon Dark Matter Collaboration (GADMC) is pursuing a multi-staged experimental program aiming to improve the sensitivity down to the ``neutrino fog''~\cite{Aalseth:2017fik}.  
Currently, GADMC is preparing for the DarkSide-20k experiment~\cite{Aalseth:2017fik} which features a 50 tonne underground argon dual-phase \TPC\ with Silicon Photomultiplier readout.

\begin{figure}[tbp]
	\centering 
	\includegraphics[width=.75\columnwidth]{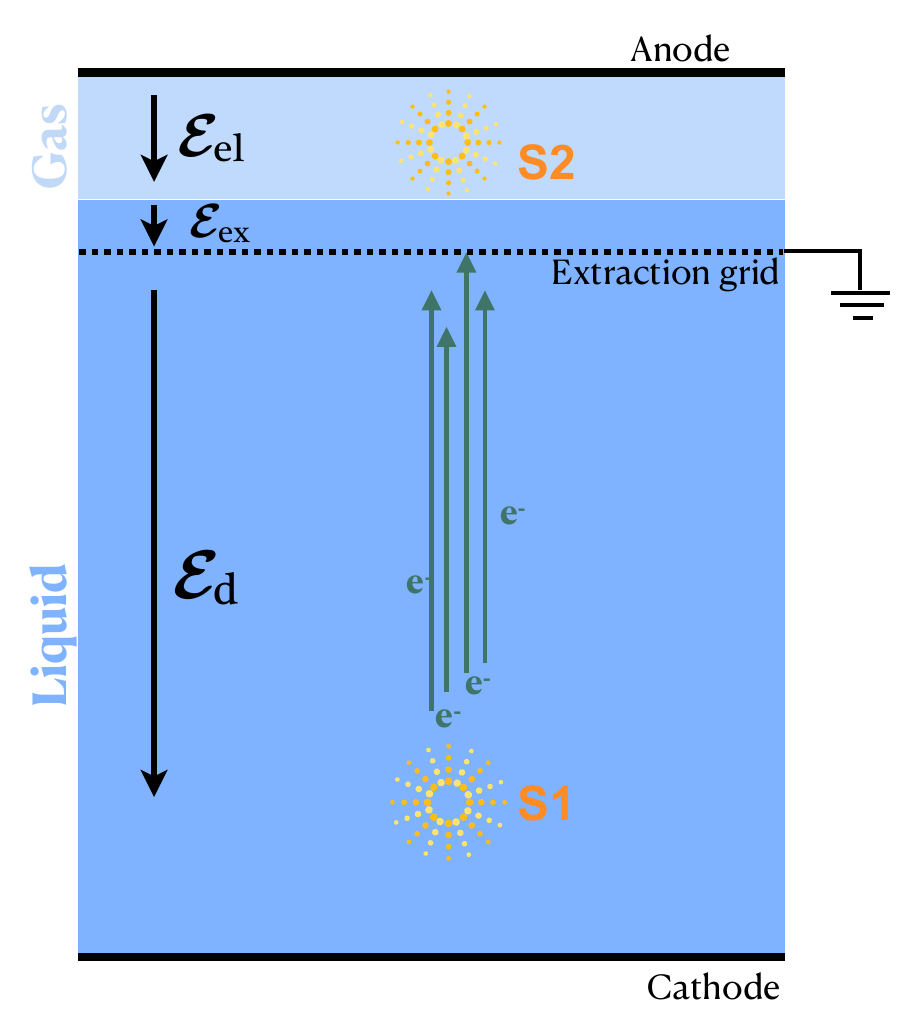}
	\caption{\label{fig:tpcworking} Conceptual sketch of the working principle of a double-phase argon \TPC, illustrating a typical event with both a primary scintillation signal (S1) and a secondary electroluminescence signal (S2). The electric field in three different regions of the \TPC\ is indicated here as the drift field (\edrift), extraction field (\eex) and electroluminescence field (\eel). Figure from Ref.~\cite{Agnes:2021zyq}.}
\end{figure}
The working principle of an argon dual-phase \TPC\ is depicted in Fig.~\ref{fig:tpcworking}. The \TPC\ contains a volume of liquid argon (LAr) with a thin layer of gaseous argon, the gas pocket, on the top. The elastic scattering of a hypothetical WIMP particle 
with an Ar nucleus in the \TPC\ would originate a NR of kinetic energy ranging from $\sim 20$ to $\sim 100$~keV,
which ionizes the medium along 
its trajectory. The energy deposition of an ionizing particle which travels in the liquid volume produces excitation and ionization, giving rise to excited argon dimers (Ar$_2^*$) and to electron-ion pairs. The de-excitation of Ar$_2^*$ dimers, some of which are produced by electron-ion recombination, emits scintillation light, which produces the S1 signal. 
The residual unrecombined ionization electrons are swept away from the interaction site and drifted towards the liquid-gas interface by an appropriate electric field, the drift field \edrift. 
They are extracted to the gas phase and accelerated by intense fields, the extraction field \eex\!  and the electrolumiscence field \eel, respectively. 
Accelerated electrons in the gas phase emit light by electroluminescence~\cite{Aalseth:2020zdm,Buzulutskov:2020xhd}, which is the S2 signal. 
The S1 and S2 signals are separated by the time interval corresponding to the electron drift time from the interaction site to the gas phase.  The S2 signal intensity is proportional to the number of extracted electrons.
The recombination of electrons with ions produces Ar dimers at the expense of free charge and therefore affects the balance between the intensity of S1 and S2 signals. 

A dual-phase \TPC\ could potentially offer a directional sensitivity for the events featuring long straight ionization tracks, thanks to the mechanism of 
columnar recombination~\cite{Jaffe:1913gs,Birks:1951boa,Cataudella:2017kcf}. 
When the track is nearly parallel to \edrift, electrons pass through the electron-ion column from the track itself and have 
a higher probability to meet an Ar ion and recombine, compared to a perpendicular track. 
Events with tracks parallel to \edrift\ are therefore expected to have an enhanced S1 and a reduced S2.
According to Refs.~\cite{Nygren:2013fy,Cao:2015ks}, the directional dependence occurs only if
 the charge cloud around the ionization track is anisotropic, namely
when the ionizing track is longer than the Onsager radius $r_O$, the distance between an ion and a free electron for
which the electrostatic potential energy equals the thermal kinetic 
energy of the electron.
As $r_O = e^2/(6 \pi \epsilon_0 \epsilon_r k_B T)$ is about \SI{80}{\nano\meter} in LAr ($T = 87$~K, $\epsilon_r = 1.5$), argon ions 
with kinetic energy above $\sim \SI{40}{\kilo\electronvolt}$ have a range longer than $r_O$~\cite{srim}. 
Therefore, an argon dual-phase \TPC\ could potentially be direction-sensitive in the energy range 
of interest for WIMP searches. 
However, calculations and simulations~\cite{Wojcik:2003ja,Wojcik:2016gy} show 
that the mean thermalization distance of electrons in LAr is about 
\SI{2.6}{\micro\meter}, which is much longer than the Onsager radius 
and of the range of WIMP-induced recoils. As recombination mostly takes 
place when electrons are fully thermalized, the directional 
sensitivity could hence be diluted by electron diffusion during thermalization. 

The SCENE Collaboration has provided a hint of directional sensitivity in the S1 signal for NRs of about 
\SI{60}{\kilo\electronvolt}~\cite{Cao:2015ks}, and specifically a difference of about 7\% on S1 for NRs parallel
and perpendicular to the drift field, at \edrift=193~V/cm. 

The breakthrough that the directional sensitivity of an argon \TPC\ would offer in 
the framework of direct dark matter searches motivated the Recoil Directionality (\ReD) experiment, 
as a part of the program of the GADMC. \ReD~\cite{Agnes:2021zyq} has been designed and performed with 
the goals to scrutinize the hint by SCENE, by testing a directional effect with a size as reported by 
SCENE, and additionally to provide new experimental data to improve the understanding of recombination and 
thermalization of electrons in LAr.

To this aim, a miniaturized argon dual-phase \TPC\ was irradiated with neutrons at INFN, Laboratori Nazionali del Sud, to produce NRs at a variety of angles with respect to the \TPC\ drift field. 
The kinetic energy of NRs is around \SI{70}{\kilo\electronvolt}, which falls in the range of interest of WIMP search in Ar and corresponds to an ion range larger than the Onsager radius.
This work is organized as follows: Sect.~\ref{sec:ArResponse} discusses the models to
describe the response of an argon dual-phase \TPC\ to NRs of the energy relevant for
dark matter searches, including the potential directional dependence. The experimental layout of \ReD\
and the description of the individual detectors are given in Sect.~\ref{sec:DetectorLayout} and \ref{sec:Detectors}, 
respectively. 
The data treatment, including reconstruction calibration, event selection, and the subsequent statistical analysis for
directional sensitivity are presented in Sect.~\ref{sec:DataAndResult} and~\ref{sec:StatisticalAnalysis}. 
The results and their potential impact are discussed in Sect.~\ref{sec:discussion}, followed by the
summary of conclusions in Sect.~\ref{sec:conclusions}.

\section{The response of Ar to nuclear recoils} \label{sec:ArResponse}
WIMPs deposit energy in LAr through elastic scattering on Ar nuclei. 
The subsequent energy loss of the \NR\ involves nuclear stopping, ionization, charge recombination, and scintillation. 
Through the series of physical processes, the total energy deposited in the \TPC\ is eventually divided into the detectable photons (S1) and electrons (S2), and the undetectable phonons (heat). 

Directional modulation of charge recombination is expected when the spatial charge distribution of ionization is anisotropic. Conventional \NR\ charge recombination models often assume an isotropic charge distribution. For example, the commonly-used Thomas-Imel model~\cite{Thomas:1987ek,Szydagis:2011tk} assumes that charges are uniformly distributed in a cubic box of size $a$; the only free parameter for a given detector material is the initial charge $Q_0$. 
The probability of charge surviving recombination under the electric drift field \edrift\ is 
\begin{equation}
    p(a,Q_0) = \frac{\mathcal{E}_d}{\xi(a,Q_0)}\ln\left(1+\frac{\xi(a,Q_0)}{\mathcal{E}_d}\right),
\end{equation}
where
\begin{equation} 
\xi(a,Q_0)= \frac{\alpha Q_0}{4a^2\mu_-}; \label{eq:TIxi}
\end{equation}
$\alpha$ is the Langevin
recombination coefficient~\cite{Langevin1903,Bubon:2016hc}, which depends on the carrier  
mobilities ($\mu_-$ and $\mu_+$ for electrons and ions, respectively) and on the dielectric constant as
\begin{equation}
\alpha = \frac{(\mu_-+\mu_+)}{\epsilon_0 \epsilon_r}. \label{eq:alpha}
\end{equation}

In order to introduce the directionality, the electron distribution after thermalization needs to be included in the model. One approach is to use the Jaff\'e  model~\cite{Jaffe:1913gs,Birks:1951boa}, commonly referred to as the columnar recombination model, where the charge distribution is modeled by a column with radius $b$, length $l$, and angle $\theta$ between its axis $\hat{r}_0$ and the drift field \edrift. The Jaff\'e model is commonly adopted for the straight tracks from minimum ionizing particles. Since \NR\ tracks are more localized, a more general and flexible parameterization
of the charge distribution $q_0(\vec{r})$ has been proposed by Cataudella et al.~\cite{Cataudella:2017kcf}, which consists of 
a three dimensional Gaussian with an elliptical profile
\begin{equation}
    q_0(\vec{r}) = \frac{Q_0}{(2\pi)^{3/2}R\sigma^3}\exp{\left(-\left(\frac{\vec{r}\cdot\hat{r}_0}{R\sigma}\right)^2 - \left(\frac{\vec{r}\times\hat{r}_0}{\sigma}\right)^2 \right) }, \label{eq:clouddist}
\end{equation}
where $\sigma$ characterizes the size of the distribution, $\hat{r}_0$ is the direction vector of the long axis, and $R$ is the aspect ratio between the long and short axes. 
 The  probability of charge surviving recombination is calculated in Ref.~\cite{Cataudella:2017kcf} as
\begin{equation}
\label{eq:rec_direction}
    p(R,\theta,Q_0) = -\frac{\mathcal{E}_d f(R,\theta)}{\xi_m}\mathrm{Li}_2\left(-\frac{\xi_m}{\mathcal{E}_df(R,\theta)}\right),
\end{equation}
where
\begin{equation}
\xi_m=\frac{\alpha Q_0}{2\pi \sigma^2\mu_-} \label{eq:ximCataudella}
\end{equation}
is the generalization of the Thomas-Imel parameter $\xi$ of
Eq.~\ref{eq:TIxi} and $\mathrm{Li}_2$ is the second order polylogarithm function.
The term $f(R,\theta)$ captures the directionality dependence and it has 
the functional form
\begin{equation}\label{eq:rec_modified_TI_ftheta}
    f(R,\theta) = \sqrt{R^2\sin^2\theta+\cos^2\theta},
\end{equation}
being $\theta$  the angle between $\hat{r}_0$ and \edrift.
When $R=1$, $f(R,\theta)=1$, so directionality vanishes and 
Eq.~\ref{eq:rec_direction} reduces to the Thomas-Imel model.

Since directionality effects do not occur before recombination, well-established models are used here to describe the S1 and S2 yields, that for \NRs\ also depend on nuclear and electronic quenching. Following the Lindhard model~\cite{Lindhard:1963vo,Bezrukov:2010qa}, the nuclear quenching factor, i.e. the ratio of the visible energy in the excitation and ionization channel to the total recoil energy, is described by 
\begin{equation}
    f_{n}(\varepsilon) = \frac{kg(\varepsilon)}{1+kg(\varepsilon)},
\end{equation}
where $k=0.133\, Z^{2/3}A^{-1/2}$ is a dimensionless factor depending on the Ar target nucleus ($A=40$, $Z=18$); the function $g(\varepsilon)$ is numerically approximated by 
Lindhard~\cite{Lindhard:1963vo} and it has the form
\begin{equation}
g(\varepsilon) = 3\,\varepsilon^{0.15}+0.7\,\varepsilon^{0.6}+\varepsilon;
\end{equation}
finally  $\varepsilon$ is the dimensionless reduced energy
\begin{equation}
    \varepsilon = \frac{4\pi \epsilon_0 a}{2e^2Z^2}E_{r} 
    = 11.5\, Z^{-7/3}\frac{E_r}{[\si{\kilo\electronvolt}]},
\end{equation}
being  $E_r$ the recoil energy  and $a$ the Thomas-Fermi screening length, calculated from
the Bohr radius $a_0$ as $a = 0.626 \cdot a_0 \cdot Z^{-1/3}$~\cite{Bezrukov:2010qa}.

The measurable energy is further reduced by electronic quenching, following 
the Mei model~\cite{Mei:2008ca}
\begin{equation}
    f_{l} = \frac{1}{1+k_e s_e}, 
\end{equation}
where $s_e=k\varepsilon^{1/2}$ is the dimensionless electronic stopping power 
and $k_e$ is associated to the original parameter $K_e$ of 
Ref.~\cite{Mei:2008ca} as $k_e = K_e (\mathrm{d}E/\mathrm{d}x)_e / (s_e \rho_m)$, with $\rho_m$ being the mass density of LAr. 

Summing up the components, the expectation of total quanta $\langle N_0 \rangle$, ionization $\langle N_\mathrm{i} \rangle$ and excitation $\langle N_\mathrm{ex} \rangle$ from a \NR\ of energy $E_r$ in LAr before recombination is
\begin{eqnarray}
    \langle N_0 \rangle &=& \frac{E_r f_n f_l}{W_\mathrm{ph}} \\
    \label{eq:NexNiYield}
    \langle N_\mathrm{i} \rangle &=& \langle N_0 \rangle \frac{1}{1+N_\mathrm{ex}/N_\mathrm{i}} \\
    \langle N_\mathrm{ex} \rangle &=& \langle N_0 \rangle - \langle N_\mathrm{i} \rangle 
\end{eqnarray}
where $W_{ph}$ is the average energy required to produce one scintillation photon in LAr and \NexNi\ is 
the excitation-to-ionization ratio directly induced by the fast ion and by its secondaries.
As a first approximation, \NexNi\ is usually treated as an energy independent constant~\cite{Doke:2002oab,Hitachi:2021hac}, which is related to the atomic levels in argon. However, the distribution of momentum transfer to electrons in the electronic stopping power is energy-dependent, which motivates the introduction of a variable \NexNi\, vs. energy. This is corroborated by the SCENE data~\cite{Cao:2015ks}, which also indicate an increase in \NexNi\ with respect to the \NR\ energy. The \NexNi\ values adopted for this work are taken from Table VIII of Ref.~\cite{Cao:2015ks}, with a
linear interpolation between the energy points. The \NexNi\ at zero energy is set to the commonly-adopted
value of $0.2$. 

The detectable electron and photon yields after recombination are
\begin{eqnarray}
\langle N_\mathrm{e^-} \rangle &=& \langle N_\mathrm{i} \rangle p(R,\theta,Q_0) = \langle N_\mathrm{0} \rangle \frac{p(R,\theta,Q_0)}{1+N_\mathrm{ex}/N_\mathrm{i}} \label{eq:ne} \\ 
\langle N_\mathrm{ph} \rangle &=& \langle N_0 \rangle -  \langle N_\mathrm{e^-}\rangle. \label{eq:nph} 
\end{eqnarray}



The capability to measure the \NR\ direction can be hidden by random fluctuations in S1 and S2. Indeed, the intrinsic fluctuations during signal generation in LAr and the detector resolution arising from signal propagation, collection, and amplification contribute to smear out the S1-S2 two-dimensional spectrum of a mono-energetic \NR. 
The intrinsic fluctuations are present for both the charge and light channels. The fluctuation in the total number of visible quanta $N_0$ is assumed here to be Gaussian distributed with a Fano factor $F=0.107$ \cite{szydagis2021review}: 
\begin{equation}
    N_0 \sim \mathrm{Gaussian}(\langle N_0 \rangle, \sqrt{F\langle N_0\rangle}).
\end{equation}
The partition of $N_0$ between $N_\mathrm{e^-}$ and $N_\mathrm{ph}$ follows a 
binomial distribution governed by \NexNi\ and by the recombination probability (see Eq.~\ref{eq:ne}):
\begin{equation}
    N_\mathrm{e^-} \sim \mathrm{Binomial}(N_0, \langle N_\mathrm{e^-} \rangle / \langle  N_0 \rangle)
\end{equation}
and $ N_\mathrm{ph} =N_0-N_\mathrm{e^-}$.

The \TPC\ signals S1 and S2 are measured in units of photo-electrons (PE) in the photosensor. The scintillation light collection efficiency and the photosensor quantum efficiency are less than unity, so that S1 has a gain $g_1 = \mathrm{S1}/N_\mathrm{ph} [\si{PE/ph}]$ less than one. Charges are amplified and converted to photons through the gas pocket electroluminescence, so that S2 has a gain $g_2 = \mathrm{S2}/N_\mathrm{e^-} [\si{PE/e^-}]$ greater than one. 

The stochastic processes of collection of the scintillation light can be described by a binomial distribution, using the gain $g_1$. For S2, the electroluminescence process is described by a Poisson distribution. The detector response also includes a position-dependent non-uniformity which could in principle be corrected in analysis. Practically, a small residual error will be present, which can be modeled by an additional Gaussian smearing of standard deviation 
$\sigma_ \mathrm{S1}^{*}$ and $\sigma_ \mathrm{S2}^{*}$ for S1 and S2, respectively. 
Approximating the S1 and S2 distributions with Gaussians, the total contribution from detector response is 
\begin{eqnarray}
    \mathrm{S1} &\sim& \mathrm{Gaussian}\left( \langle N_\mathrm{ph} \rangle g_1, \sqrt{\langle N_\mathrm{ph} \rangle g_1 (1-g_1) + \sigma_ \mathrm{S1}^{*2}}\right) \\
    \mathrm{S2} &\sim& \mathrm{Gaussian}\left( \langle N_\mathrm{e^-} \rangle g_2, \sqrt{\langle N_\mathrm{e^-} \rangle g_2 + \sigma_\mathrm{S2}^{*2}} \right).
\end{eqnarray}

In conclusion, the argon dual-phase \TPC\ response to a mono-energetic \NR\ follows the probability density function coming from the convolution of the detector and physical terms:
\begin{eqnarray}
    P(\mathrm{S1},\mathrm{S2}) & = &
    P_\mathrm{detector}(\mathrm{S1}/g_1,\mathrm{S2}/g_2;N_\mathrm{ph},N_{e^-}) \nonumber\\
    & & \otimes 
    P_\mathrm{NR}(N_\mathrm{ph},N_{e^-};E_r,R,\theta)  \nonumber\\
    & = & \frac{1}{2\pi\sigma_\mathrm{S1}\sigma_\mathrm{S2}/g_1 g_2} \nonumber\\
    & &
    e^{-\frac{(\mathrm{S1}/g_1 - N_\mathrm{ph})^2}{2(\sigma_\mathrm{S1}/g_1)^2}
    -\frac{(\mathrm{S2}/g_2 - N_{e^-})^2}{2(\sigma_\mathrm{S2}/g_2)^2}} \nonumber\\
    & & \otimes\frac{1}{2\pi\sqrt{F\langle N_\mathrm{ph}\rangle\langle N_{e^-}\rangle}} \nonumber\\
    & & e^{ -\frac{(N_{e^-}+N_\mathrm{ph}-\langle N_0\rangle)^2}{2F\langle N_0 \rangle}
    -\frac{(N_{e^-}\langle N_\mathrm{ph} \rangle - N_\mathrm{ph}\langle N_{e^-}\rangle)^2}
    {2\langle N_{e^-} \rangle\langle N_\mathrm{ph} \rangle\langle N_0\rangle}}. \nonumber\\
    \label{eq:TPCResponse}
\end{eqnarray}
Later in Sect.~\ref{sec:StatisticalAnalysis}, a likelihood function is evaluated from the \TPC\ data using this probability density function. An unbinned profile likelihood study is then performed to determine the confidence interval of the directionality parameter $R$.

\section{Experimental setup} \label{sec:DetectorLayout}
The experimental layout is conceived in order to produce and detect Ar nuclear recoils of known energy and direction, by neutron elastic scattering. 
Neutrons are produced by the primary reaction p($^{7}$Li,$^{7}$Be)n, by shooting a $^7$Li beam  on a polyethylene (CH$_2$) target. 
The neutron energy $E_n$ and its direction are 
kinematically determined by measuring the energy and direction of the accompanying $^7$Be nucleus.
The neutron can undergo elastic scattering $(n,n')$ with an Ar nucleus inside the \TPC, thus producing a \NR\ 
and a secondary neutron whose energies and momenta are again correlated by two-body kinematics. The scattered neutron is
eventually detected by a neutron spectrometer made by an array of liquid scintillator (LSci) detectors; the detection of
the neutron by a specific LSci determines the energy and the direction of the Ar recoil.

\begin{figure}[tpb]
\centering
    \includegraphics[width=\linewidth]{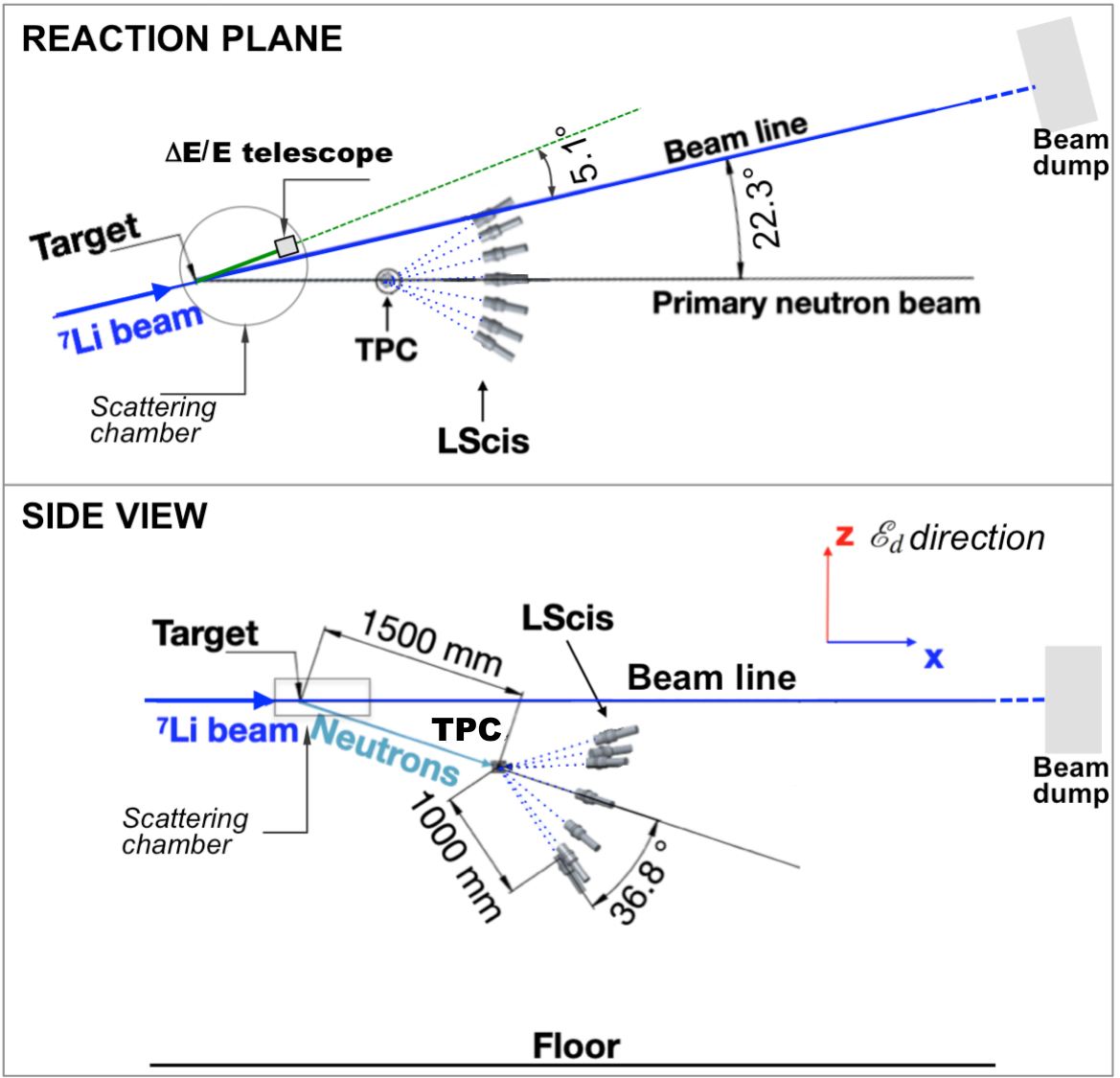}
    \caption{Schematic layout of the \ReD\ experimental setup (not in scale). Upper panel is the view of the p($^{7}$Li,$^{7}$Be)n reaction plane, lower panel is the side view.
    The primary \Li\ beam travels along the $x$ axis and enters the vacuum scattering chamber which hosts the CH$_2$ target and the $\Delta E$/$E$ telescope. 
    Neutrons emitted by the p($^{7}$Li,$^{7}$Be)n reaction undergo elastic scattering inside the \TPC\ and are eventually
    detected by one of the LScis of the neutron spectrometer, that are deployed within a cone of
    opening $\theta_{lsci}=36.8$\textdegree\ with respect to the target-\TPC\ axis. See text for more details.}
    \label{fig:schema}
\end{figure}

The conceptual layout of ReD is sketched in Fig.\,\ref{fig:schema}. The experiment deploys three detector
systems: (1) a $\Delta E$/$E$ telescope made by Si detectors, to identify \Be\ nuclei associated with 
neutrons; (2) the \TPC\ to detect the Ar NRs; (3) a neutron spectrometer made by 7~LSci detectors to detect the
neutrons scattered off Ar. The detectors of the neutron spectrometer are placed along the base circumference of a cone with axis corresponding to the target-TPC line (i.e. the direction of the incoming neutron), vertex on the TPC center and  opening angle $\theta_{lsci}$. Therefore, all LScis detect neutrons which undergo elastic scattering on Ar at the same angle and hence produce NRs of the same energy \Er. 
While the NRs tagged by the seven individual LScis all have the same energy $E_r$, 
their momenta $\vec{p}_r$ form a different angle $\theta_r$ with respect to the \TPC\ electric field ($z$ axis in Fig.\,\ref{fig:schema}), as
required to test the directional effect. As it is important for this work 
to test the response to NRs also at $\theta_r =180$\textdegree, the \TPC\ 
is placed at a different level with respect to the target, such to provide 
the incoming neutron with a momentum component along the field direction.

Once  the angle $\theta_{tpc}$ between the primary \Li\ beam direction and the target-TPC direction and the angle 
$\theta_{lsci}$ are fixed by the setup geometry, \ReD\ is tuned to select mono-energetic Ar recoils of energy \Er\ by the triple 
coincidence between the Si telescope, the \TPC\ and the neutron spectrometer. 
The operational parameters chosen for ReD are
$\theta_{tpc} = 22.3$\textdegree\ and $\theta_{lsci} = 36.8$\textdegree. The target-TPC distance and the TPC-LSci distance are 150 and
100~cm respectively, as a reasonable compromise between angular resolution and solid angle coverage: in both cases
the uncertainty on the neutron direction is driven by the dimensions of the \TPC\ and of the LSci, i.e. by the uncertainty
on the interaction point within them. Keeping the geometry fixed, the energy \Er\ of the NR can be changed by varying the primary beam energy.
The \ReD\ experimental layout was designed to allow for the measurements of NRs in the range of interest for dark matter direct searches,
between 20 and 100~keV: this can be achieved by varying the energy of the primary \Li\ beam between 20 and 34~MeV.

\subsection{\Li\ beam and target} \label{sec:nbeam}
 The primary \Li\ beam  is produced by the 15~MV TANDEM accelerator of the INFN LNS~\cite{CIAVOLA199364} at an energy of  28~MeV. The TANDEM offers an excellent resolution in the delivered energy, which is about 1\% FWHM in our case. The data reported in this work were collected 
between January 31$^{st}$ and February 14$^{th}$, 2020. The current of the \Li\ beam ranged between 5 and 15~nA,
 corresponding to $1-3 \cdot 10^{10}$~(\Li/s). 
The beam is driven to a vacuum scattering chamber, which hosts the CH$_2$ target and the $\Delta E$/$E$ telescope. Upstream the target,  the \Li\ beam is collimated to obtain a spot of 2~mm diameter at the target position. Neutrons are produced via the p($^{7}$Li,$^{7}$Be)n reaction. 
 The $\Delta E$/$E$ telescope detects the $^7$Be accompanying the neutrons that travel towards the TPC. As the accelerator 
does not allow the production of a pulsed beam,
the direct detection of \Be\ represents the best solution for event-by-event neutron tagging. The requirement to
detect \Be\ drives the choice of inverse kinematics (i.e. \Li\ beam on a hydrogenous target)~\cite{Drosg1981,Dave1982}, instead of the
direct kinematics approach (proton beam on a \Li\ target) employed by other experiments, as SCENE~\cite{Cao:2015ks}. 
 
The targets of CH$_2$ have thickness ranging between 150 and 350~\textmu g/cm$^2$, which is thin enough to allow for the escape of \Be. Due to aging effects, each target was used for about 12~hours of data taking, before being replaced by means of a 
12-target holder system placed inside the vacuum scattering chamber.

After the target, the \Li\ beam travels straight forward towards a Ta beam dump placed 3~m downstream (see 
Fig.~\ref{fig:schema}). Such a long distance is functional to minimize the background on the ReD setup due to the 
beam interaction on the beam dump.
The beam intensity was precisely measured every few hours of operation by a 
Faraday Cup deployed about 30~cm downstream the target. However, the Faraday Cup was removed during the data taking, in order
to reduce the background radiation close to the \TPC. The continuous monitoring of the beam intensity
was performed by measuring the rate of the \Li\ elastic scattering on a dedicated Si detector (not shown in Fig. \ref{fig:schema}) placed at
$\theta = 7$\textdegree\ with respect to the beam line, where no \Be\ is allowed by kinematics. 

\section{The detectors} \label{sec:Detectors}
\subsection{The $\Delta E$/$E$ telescope} \label{sec:deltaee}
Neutrons directed towards the \TPC\  
are produced in association with \Be\ nuclei of energy 
$E_{Be} = 19.0$~MeV and emitted at angle $\theta_{Be} = 5.1$\textdegree. $^{7}$Be is detected by a dedicated $\Delta E$/$E$ 
telescope placed in the scattering chamber at a distance of \SI{46}{cm} from the CH$_2$ target. The telescope is made of two Si detectors manufactured by ORTEC, having thickness of 20~\textmu m and 1000~\textmu m, respectively; the $^{7}$Be loses about 7.6~MeV crossing the thinner stage and it is stopped in the thicker one. The detectors have a 100\% efficiency for light charged particles detection and energy resolution of  about 1\%. The telescope is collimated using an Al shield with a hole of 2~mm diameter.  For the fine tuning of the position, the telescope holder is mounted on a two axis remotely-controlled stepper motor which can operate in vacuum.
The detectors are 
readout from a standard spectroscopic chain made by a pre-amplifier 
and a charge-sensitive amplifier, 
with 1~\textmu s  shaping time. 
%

The combined measurement of $\Delta E$ and $E$ provides the discrimination in $Z$, which is necessary to distinguish the
interesting Be from the far more abundant elastically-scattered Li. 
%
%
%
%
\begin{figure}
	\centering
	\includegraphics[width=0.95\columnwidth]{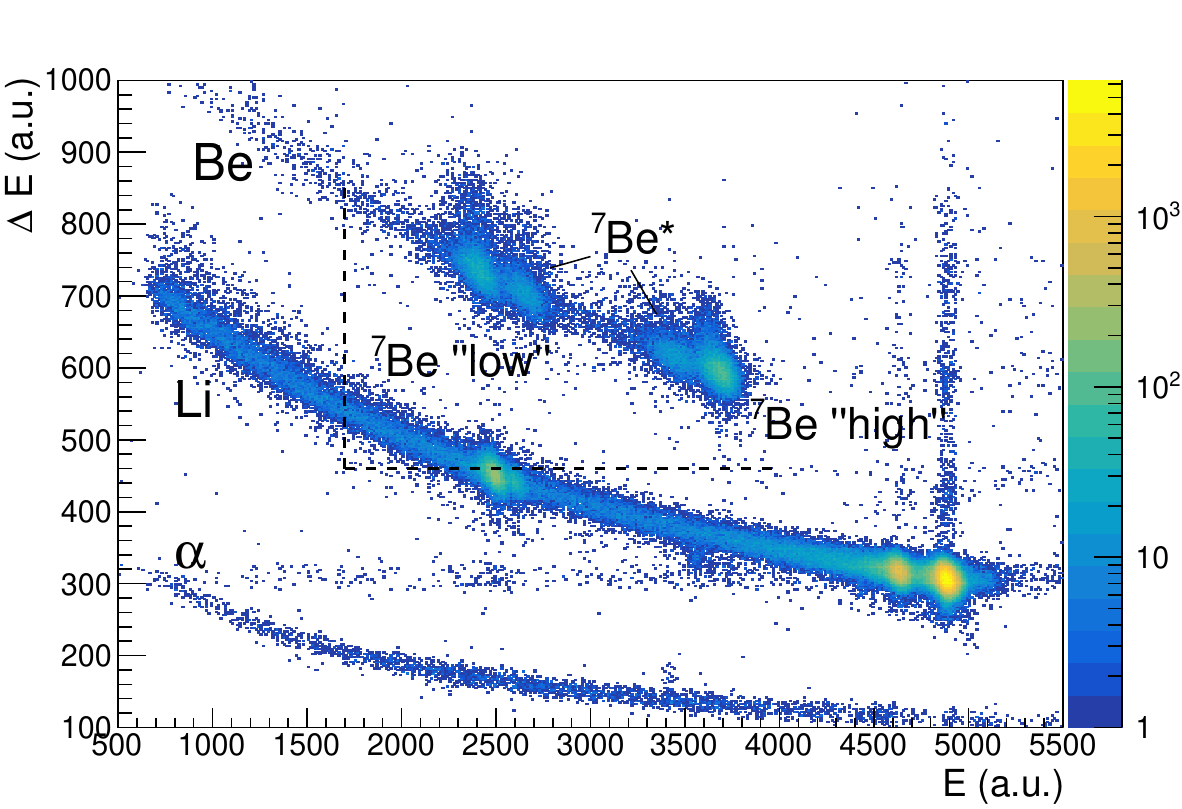}
	\caption{$\Delta E$ vs. $E$ scatter plot obtained from the irradiation of a CH$_2$ target with a 28-MeV $^7$Li beam. The bands 
		identify nuclei of different $Z$ ($\alpha$, Li and Be), as discussed in the text. Neutrons traveling towards the \TPC\ are 
		produced in association with the $^7$Be nuclei of the \emph{locus} labeled as ``$^7$Be low''. The dashed lines show the 
		thresholds used  in the normal operating conditions for the $\Delta E$ and $E$ detectors, and which are meant to suppress the 
		dominant contribution from Li. } \label{fig:banana1}
\end{figure}
Fig.~\ref{fig:banana1} shows the $\Delta E$ vs. $E$ scatter plot, upon the irradiation of the CH$_2$ target with the $^7$Li beam.
The central, and most intense, band is created by Li ($Z=3$), mostly by elastic scattering on H and C. The uppermost band is due to Be ($Z=4$). 
As the reaction p($^{7}$Li,$^{7}$Be)n occurs in inverse kinematics, two different solutions at the same angle
$\theta_{Be} = 5.1$\textdegree\ are allowed, with \Be\ having energy of 19.0~MeV (``low energy'') and 20.4~MeV (``high energy''), respectively. 
Neutrons in association with the ``low energy'' \Be\ are those travelling towards the \TPC\ ($\theta_{n} = 22.3$\textdegree), with $E_n = 7.3$~MeV
kinetic energy. The
``high energy'' \Be\ is associated with neutrons of $E_n = 2.7$~MeV emitted at $\theta_n = 44$\textdegree: 
these neutrons do not hit directly the \TPC, but can contribute to accidental coincidences due to scattering on the floor or on the walls. 
In  Fig.~\ref{fig:banana1}  the \emph{loci} from the two $^7$Be solutions are visible and clearly separated; the population between them is due to the inelastic interaction p($^{7}$Li,$^{7}$Be*)n', which also emits a neutron. Because of the finite extension of the beam spot 
and of the beam angular divergence, neutrons associated with the $^{7}$Be* detected at $\theta_{Be}$ can still travel inside 
the \TPC\ and produce an interaction; they also contribute to the diffuse 
background, e.g. upon scattering on the walls or on the floor of the 
experimental area.

\begin{figure}
	\centering
	\includegraphics[width=0.95\columnwidth]{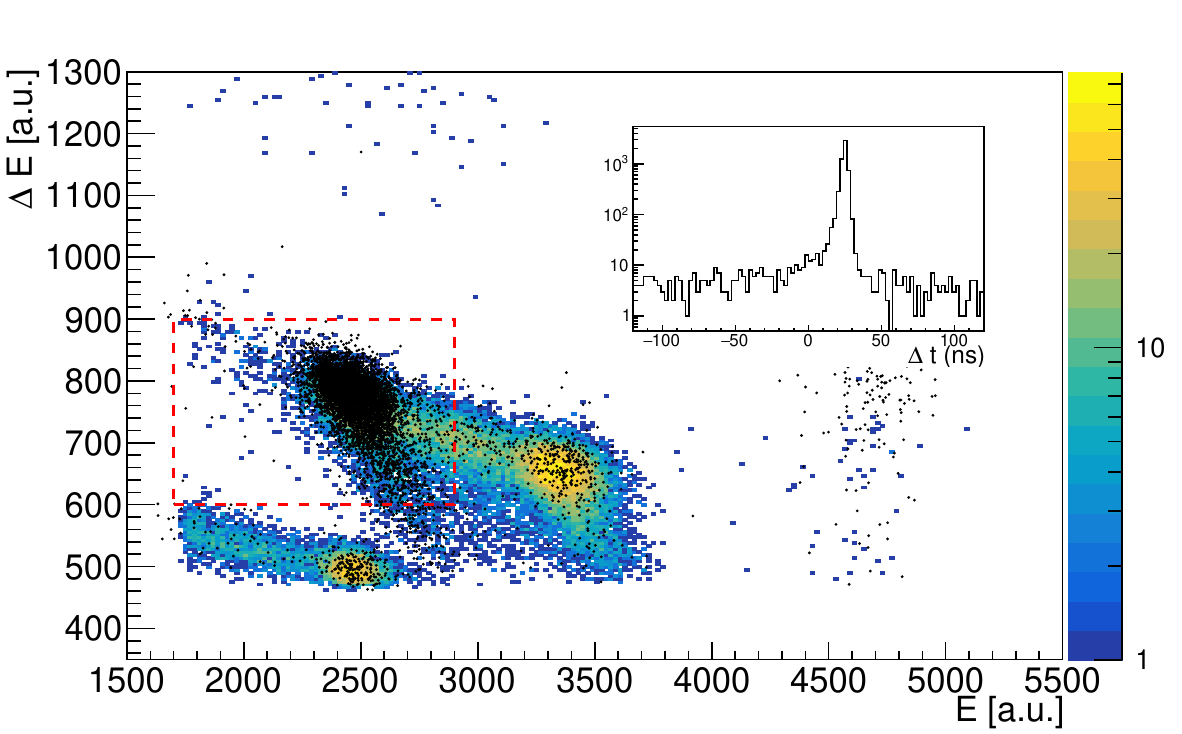}
	\caption{$\Delta E$ vs. $E$ distribution obtained from the irradiation of a CH$_2$ target with a 28-MeV $^7$Li beam (color scale). 
		The black dots are the events detected by the Si telescope in coincidence (within 200~ns) with an S1 signal in the \TPC\ having a 
		time profile compatible with a neutron-induced interaction. The dashed red box represents the \Be\ selection cut used in 
		the following analysis and described in Sect.~\ref{sec:EventSelection}. Inset: distribution of the time difference $\Delta t$ between 
		\TPC\ and Si telescope for events within the 200~ns coincidence gate.} \label{fig:banana2}
\end{figure}

 In order to suppress the dominant contribution from \Li\ elastic scattering, the thresholds for the $\Delta E$ and $E$  detectors, shown in Fig.~\ref{fig:banana1} as dashed lines, are used during the data acquisition. Fig.~\ref{fig:banana2} displays the $\Delta E$ vs. $E$ scatter plot, acquired with the thresholds 
of Fig.~\ref{fig:banana1}, without (color) and with (dots) the requirement of coincidence with an event in the \TPC\ compatible with 
a neutron interaction and within a 200~ns gate. As expected, neutron events in the \TPC\ are mostly associated with a ``low-energy'' \Be\
nucleus detected by the Si telescope. The dashed red box represents the \Be\ selection cut used in the 
following analysis and described in Sect.~\ref{sec:EventSelection}. 
%

\subsection{The Time Projection Chamber} \label{sec:2a}
The heart of the ReD system is the dual-phase Ar \TPC, whose detailed description and 
performance are reported in~\cite{Agnes:2021zyq}. It is a cubic volume of $5 \times 5 \times 6$ 
cm$^3$, delimited on the side walls by acrylic plates interleaved with 
specular reflector foils, and the top and bottom by two transparent acrylic windows. 
The top and bottom windows are coated with a thin transparent conductive layer (indium-tin oxide, ITO), so 
they can be given an electric potential and be operated as anode and cathode, respectively. The extraction grid is a stainless 
steel mesh, having 95\% optical transparency; it is located 10~mm below the anode window and it is kept electrically grounded.
All internal surfaces are coated with a wavelength shifter (tetraphenyl-butadiene, TPB): it converts the UV light emitted by Ar scintillation (128~nm) into visible
light, which better matches the sensitivity of typical photosensors. The lower part of the \TPC\ contains LAr:  
the liquid fills the entire volume between the cathode and the extraction grid, plus 3~mm above the grid. The gas pocket 
is produced by means of a heater and it occupies the 7-mm thick region between the liquid surface and the anode. 

The \TPC\ electric fields which are set for this work are: drift field (\edrift) of 152~V/cm; extraction field (\eex) of 3.9~kV/cm; and electroluminescence field (\eel) of 5.9~kV/cm.
The maximum drift time is about 66 \textmu s: this is the time required for an electron produced at the
cathode to travel until the liquid surface. Due to a continuous recirculation loop of the liquid through a SAES getter,
the purity of argon is such that the electron life time before 
capture by electronegative impurities is $> 1$~ms, i.e. much longer than the 66-\textmu s maximum drift time~\cite{Agnes:2021zyq}. 
The extraction field is strong enough to give a 100\% extraction efficiency of the electrons from the liquid to the 
gas phase~\cite{Chepel:2012sj}.

After the UV photons from scintillation and electroluminescence of Ar are shifted to the visible range by the TPB coating, they 
can be detected by customised NUV-HD-Cryo Silicon PhotoMultipliers (\SiPMs) from Fondazione Bruno Kessler, which can be operated at cryogenic 
temperature~\cite{Gola:2019idb}. The \SiPMs\ are assembled in two $5 \times 5$~cm$^2$ tiles, each containing 24 devices of 
dimensions 11.7~mm$\times$7.0~mm and arranged in a $4 \times 6$ array.
The tiles are placed behind the top and bottom acrylic windows of the \TPC, providing a 30\% 
total optical coverage. As the position of the S2 event in the gas phase can be used to estimate the \xy\ coordinate of the 
original interaction point in the \TPC, the \SiPMs\ of top tile are readout in 22 channels for improved resolution: 20 \SiPM\
are readout individually, while 4 lateral \SiPMs\ are summed in pairs and grouped into two readout channels. 
The \SiPMs\ of the bottom tiles are summed in groups of twelve, hence giving two readout channels.
Two custom-made Front-End Boards (FEB) 
supply power to the \SiPMs\ and amplify the output signals at cryogenic temperature. The \SiPMs\ are operated at \SI{+7}{V} of overvoltage 
with respect to the breakdown voltage. Due to the presence of resistors in the bias chain, the effective 
overvoltage of the \SiPMs\ gets smaller than the nominal \SI{+7}{V} when the bias current of the devices is high. 
This typically happens when the \SiPMs\ are exposed to a significant amount of light, e.g. 
due to the high interaction rate under beam irradiation, and causes a change in the \SiPM\ response 
(see Sect.~\ref{sec:Calibration}). 

More details about the cryogenic setup, the \TPC, the photosensors and the readout system can be found in~\cite{Agnes:2021zyq}.

\subsection{The neutron spectrometer} \label{sec:spectrometer}
The neutron spectrometer used in \ReD\ is made of seven 3-inch liquid scintillator (LSci) cells, individually read-out by
photomultipliers (PMTs). The assembly includes the liquid scintillator cell, a ETL-9821B PMT and the front-end electronics with the 
amplifier. The cells are filled with the EJ-309 liquid scintillator by Eljen Technologies, which features a very powerful
neutron-$\gamma$ discrimination based on the time pattern of the scintillation pulse. 

The neutron detection efficiency of the detectors was measured individually by using a $^{252}$Cf 
source~\cite{Stevanato2014,simophdthesis} and found to be about 28\% for the 7-MeV neutrons of interest 
for this work. 
The calibration of the energy scale was performed with $\gamma$-ray sources 
($^{241}$Am, $^{137}$Cs and $^{22}$Na). Dedicated measurements taken with the annihilation $\gamma$-rays from the 
$^{22}$Na source confirmed the time resolution to be better than 1~ns (rms).

The scintillators  identify Ar recoils of the same energy but different angles $\theta_r$ with respect to the \TPC\ drift field \edrift:  
$\theta_r$=180\textdegree 
(one LSci), 90\textdegree (two LScis, read out individually and labeled as
``90\textdegree $l$'' and ``90\textdegree $r$''), 40\textdegree (two LScis, summed) and 
20\textdegree (two LScis, summed).

\subsection{Data acquisition and control infrastructure} 
The output signals from all of the detectors are sent to CAEN V1730 Flash ADC Waveform Digitizers and 
digitized with 14-bit resolution at a sampling rate of 500 MHz. In total a signal of 100~\textmu s (50k samples) is 
acquired at each trigger: this is sufficiently long to contain the S1 and S2 signals of the \TPC, given the maximum 
drift time of 66~\textmu s for events occurring close to the cathode. About 10\% of the digitization window is reserved 
for the pre-trigger. 
Two 16-channel CAEN V1730 boards were used for the measurement, synchronized with a daisy chain. 

The data acquisition (DAQ) software was built upon a package 
developed for the PADME experiment~\cite{Leonardi_2017} and based on the CAEN Digitizer Libraries. 
The trigger logic is implemented by means of an external NIM logic module 
as:
\begin{equation}
\texttt{SiTel} \land (\texttt{TPC} \lor \texttt{LScis})
\end{equation} 
where: \texttt{SiTel}, \texttt{TPC} and \texttt{LScis} are the trigger signals from the Si telescope, the \TPC\ and 
the neutron spectrometer, respectively.
The Si telescope trigger (\texttt{SiTel}) is built as the coincidence of the $\Delta E$ and $E$ 
detectors, with the thresholds displayed in Fig.~\ref{fig:banana1}.
The \TPC\ trigger (\texttt{TPC}) consists in the logical \texttt{AND} between the
two readout channels of the bottom tile within a coincidence gate of 200~ns, in order to suppress the dark rate~\cite{Agnes:2021zyq}.
The individual thresholds are set to approximately 2~PE. The \TPC\ is expected to trigger with 100\% efficiency on
S1 signals from the $E_r = 72$~keV NR events ($\textrm{S1} \sim 190$~PE) which are of interest for this work, although trigger
inefficiencies can possibly come from pile-up. 
Finally, the neutron spectrometer trigger (\texttt{LScis}) is produced by the logical \texttt{OR} of the five readout channels 
of the seven scintillators. The energy threshold of each cell is set to approximately 20~keV$_{ee}$ (electron equivalent), 
which corresponds to about 200~keV for a proton recoil~\cite{Stevanato2014}.
This is sufficient to have a nearly-100\% trigger efficiency for the neutron events of interest, as their
elastic scattering on the scintillator produces protons of average energy $\sim 3.6$~MeV, giving a 1.1~MeV$_{ee}$ 
signal~\cite{Stevanato2014}.


All detectors and sensors of the setup can be operated and read out remotely by means of a slow control system made of a suite 
of LabVIEW-based~\cite{LabView} applications. All parameters under control (e.g. temperatures, bias voltages, leakage currents) 
are monitored continuously, and readings are stored in a database every 10~s.

\section{Event processing and selection} \label{sec:DataAndResult}
\subsection{Event reconstruction and calibrations} \label{sec:Calibration}

The raw data from the \TPC\ are the digitized waveforms of each of the \SiPM\ channels, from which the event type, time, and 3D 
position were reconstructed following the procedure described in~\cite{Agnes:2021zyq}. The waveforms of all \SiPM\ channels
were baseline-subtracted, equalized according to the individual gains and summed. The summed waveform was filtered
with a \SI{20}{\nano\second} moving average window and then scanned by a dedicated pulse-finder algorithm, to search
for possible S1 and S2 signals. Each pulse was classified as
either S1 or S2 by using the pulse shape parameter \fprompt, defined as the ratio of the charge in the first
\SI{700}{\nano\second} over the total charge: pulses with $f_\mathrm{p}<0.2$ are classified as S2. The total charge was
then normalized according to the Single Electron Response (SER) of each \SiPM\ channel, so to provide the S1 and S2
in units of PE. The pulse-finder algorithm is fully efficient for S1 signals above a few keV.  The time delay between the S1 and S2 pulses, i.e. the electron drift
time \tdrift, was used to estimate the $z$ coordinate of the interaction below the liquid-gas interface. 
Events with a single S1 pulse and a S2 pulse with $t_\mathrm{drift}$ between $\SI{6}{\micro\second}$ and 
the maximum drift time were kept for the subsequent analysis.
The cut $t_\mathrm{drift} > \SI{6}{\micro\second}$
removes the events produced just below the extraction grid of the \TPC, in which the S1 and S2 pulses are 
piled-up.
The approximate \xy\ position of the event was evaluated as the charge-weighted center of the S2 signal 
in the top \SiPM\ array. The parameter \fprompt\ defined above was also used to 
perform the NR/ER discrimination: S1 pulses with $f_\mathrm{p} > 0.4$ are selected as from \NR. This simple
cut was shown to allow for a NR/ER separation better than $2~\sigma$ for S1 above 50~PE~\cite{Agnes:2021zyq}. 

The SER and the duplication factor $K_\mathrm{dup}$ were studied by irradiating the \SiPMs\ with a 403-nm laser source and by modeling the 
photon counting statistics according to the Vinogradov distribution~\cite{vinogradov2009}. 
The calibration was performed channel by channel, as described in \cite{Agnes:2021zyq}.
Typical values of $K_\mathrm{dup}$, which is the average number of secondary PEs produced by cross-talk and afterpulsing 
by each genuine primary photon on the \SiPM, are between 0.31 and 0.37. The PE gain is corrected to remove cross-talk and afterpulsing according to the ratio $\frac{1}{1+K_\mathrm{dup}}$. Dedicated laser calibrations were taken every
12~hours throughout the beam time to monitor the stability of the \SiPMs.

As mentioned in Sect.~\ref{sec:2a}, the voltage drop in the bias resistor chain causes a reduction 
of the bias voltage of the \SiPMs, which is proportional to the bias current and must be properly accounted 
for in the data analysis. The bias current registered during the 
laser calibrations by the slow control system was $< \SI{0.5}{\micro\ampere}$.  During the beam irradiation,
because of the much higher interaction 
rate and the much higher amount of light hitting the \SiPMs, the bias current ranged up to 
90 (150)~\textmu A for the bottom (top) \SiPMs, depending on the intensity of the 
primary $^7$Li beam, which was not constant in time. To derive the corrections to the SER for 
each individual \SiPM, three dedicated laser runs in which the \TPC\ was 
simultaneously irradiated with high-activity radioactive sources were performed. The typical correction is of the order of $0.5\% \cdot I$, where $I$ is the bias current 
in \textmu A. 
For this reason, the SER and $K_\mathrm{dup}$ correction was 
time-dependent and calculated using the closest reading of the bias current registered by
the slow control. 
Besides the SER and $K_{dup}$, the photon detection efficiency also changes with bias voltage. A set of runs 
with \ISO{Am}{241} \SI{60}{\kilo\electronvolt} $\gamma$ and the \ISO{Li}{7} beam irradiation was performed 
to calibrate the additional bias current dependency in PE yield after the SER 
and $K_\mathrm{dup}$ correction.

The performance of the \TPC\ was characterized prior to the irradiation, 
in a dedicated campaign~\cite{Agnes:2021zyq}. Specifically, the 
scintillation gain and ionization amplification of the \TPC\ were measured 
to be $g_1 = (0.194 \pm 0.013)$~PE/photon and 
$g_2 = (20.0 \pm 0.9)$~PE/electron, respectively.  
Additional calibrations with \Am\ were taken daily during the campaign 
at LNS. These measurements confirm a light yield of $(8.53 \pm 0.19)$~PE/keV 
at 60~keV and at \edrift=150~V; this is very well consistent with the expectation of
8.6~PE/keV based on the parametrization obtained in the pre-irradiation campaign~\cite{Agnes:2021zyq}.

The daily calibration runs with \Am\ were used to evaluate the dependence of the \TPC\
response on the interaction position, and to determine the correction
factors for S1 and S2, to be later applied to the physics runs. The events featuring one single S1 and one single S2 and
having S1 compatible with the full energy deposition of the 60~keV $\gamma$-ray from \Am\ were grouped 
in a $22 \times 11$ mesh, according to the interaction position in the \TPC. The mesh has 22 entries in \xy, based on the
top SiPM channel detecting the largest fraction of the S2 signal, and 11 bins in $z$, equally spaced between
$t_\mathrm{drift}=\SI{6}{\micro\second}$ and $\SI{72}{\micro\second}$.

\begin{figure}
    \centering
    \includegraphics[width=\linewidth]{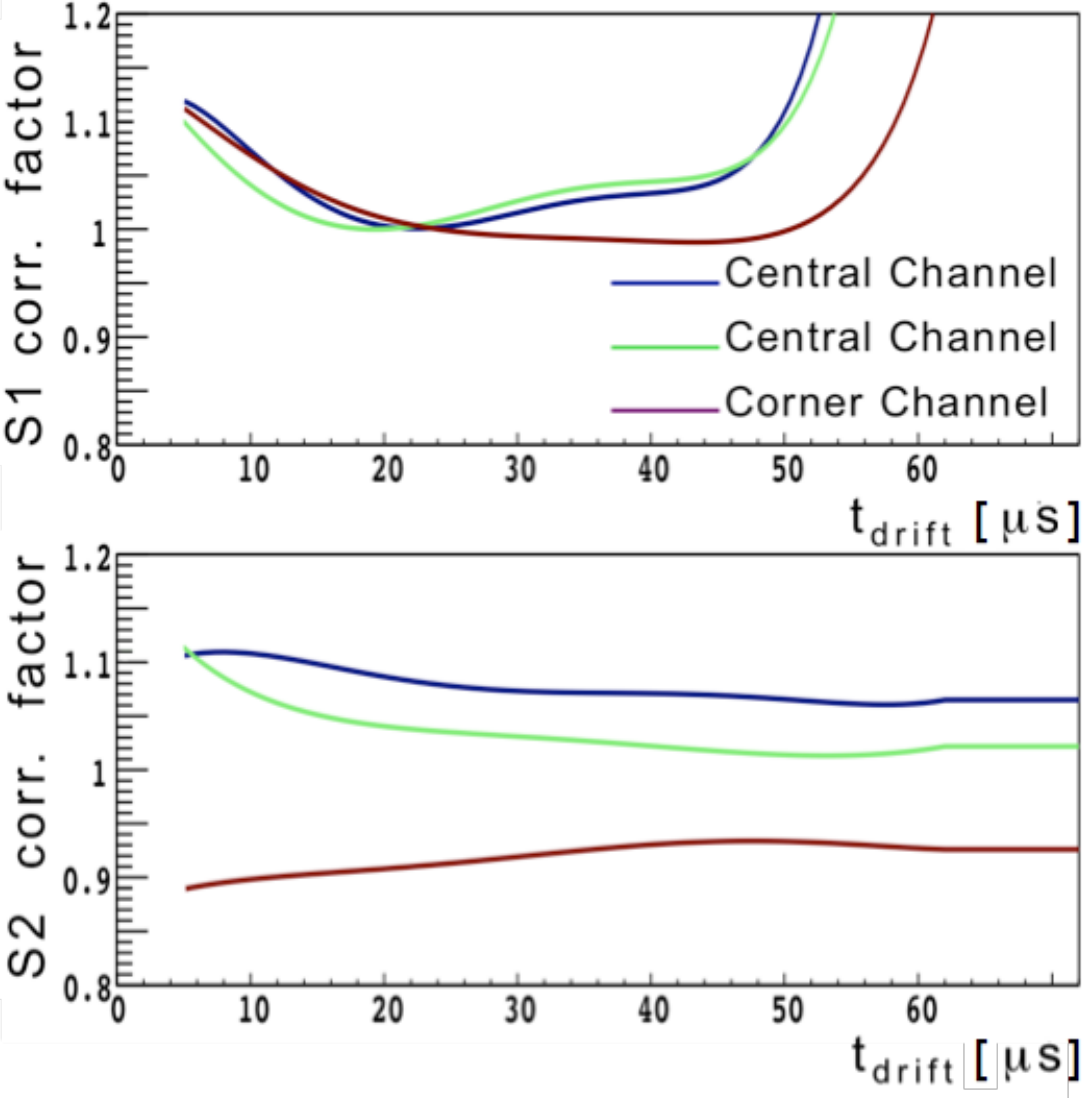}
    \caption{Examples of correction factors for the $z$ position dependence for events located below two central \SiPMs\ (blue and green)
    and below one corner \SiPM\ (red). Upper (lower) panel: correction factor for S1 (S2).}
    \label{fig:S1S2CorrExample}
\end{figure}

Firstly, S2 was corrected to account for the presence of impurities in LAr, which can cause the absorption of
electrons during their drift path. The electron life time was
estimated with an exponential fit of the S2 vs. \tdrift\ profile, restricted to the events in the  
central eight \xy\ bins. 
The $z$ dependency of S1 and S2 was further corrected by using a set of 5th-order polynomials
S1$_{i}(t_\mathrm{drift})$ and S2$_{i}(t_\mathrm{drift})$:
they are calculated by 
interpolating over the $z$-points within each bin $i$ in \xy. Three examples are shown in
Fig.~\ref{fig:S1S2CorrExample}: the correction vs. $z$ is within 10-15\%, for both S1 and S2.
No significant variation in the position correction was found throughout the sequence calibration
runs. Position dependencies mostly result from non-uniformity in the light collection efficiency within
the \TPC: 
as a consequence, the same corrections for S1 and S2 derived from \Am\ (ER) events were also applied to \NR\ events.

A simpler processing was performed for the digitized waveforms from the liquid scintillators and from the Si detectors
of the telescope. The signal in the LSci detectors was processed by calculating the total charge, integrated within
a gate of 600~ns. The ratio between the charge in the first 80~ns and the total was used as
the discrimination parameter, resulting in a neutron-$\gamma$ discrimination better than $3 \sigma$ above 
200~keV$_{ee}$~\cite{simophdthesis}. The signals from the $E$ and $\Delta E$ detectors of the telescope were
evaluated by taking the maximum of the digitized shaped waveforms from the charge-sensitive amplifier. 

The time signal of all three kinds of detectors in the setup is critical for the coincidence event selection. The time
stamp of a \TPC\ event was defined as the zero-crossing time of the pulse obtained by passing the S1 pulse through a constant 
differential filter (CDF). 
The $\Delta E/E$ telescope generates two time stamps, one for the $\Delta E$ detector and one for the $E$ detector, which were
both evaluated with CDFs. The reference time for the $\Delta E/E$ telescope used for the coincidence was taken as 
the average of the two time stamps. Finally, the time stamp for the neutron spectrometer was defined as the zero-crossing CDF
time of the digitized waveforms. 

\subsection{Selection of signal events} \label{sec:EventSelection}

The events of interest are triple coincidences between a \Be\ nucleus 
detected in the $\Delta E$/$E$ telescope, and the two subsequent neutron scatterings
in the \TPC\ and in the neutron spectrometer.

A clean sample of signal events with the proper topology was selected through
a sequence of cuts.
Firstly, unambiguous \TPC\ events were selected
according to same criteria of Sect.~\ref{sec:Calibration}:
events with only one S1 and only one S2, separated by a $t_\mathrm{drift}$ within
the range $[6,66]~\si{\us}$. An additional S2 ``echo'' 
signal, namely a secondary event due to photo-ionization of the 
cathode from the main S2 electroluminescence, is allowed in the time 
window $[67.5, 72]~\si{\us}$ after the primary S2.

Afterwards, events in the \TPC\ were selected by requesting that S1 is in time coincindence 
within a gate of 200~ns with the $\Delta E$/$E$ telescope and with one 
single LSci detector of the neutron spectrometer. In addition, 
neutron-induced (n,n') events in the neutron spectrometer were efficiently 
selected by PSD against the dominant $\gamma$-ray background.
The PSD based on the S1 signal of the \TPC\ was not applied. This was 
meant to avoid an undesirable S1-dependent selection efficiency, given 
the fact that the discrimination based on \fprompt\ gets progressively worse 
for S1 signals below \SI{100}{PE}.



The \Be\ ion which accompanies the neutron traveling towards the \TPC\ was selected by a combined 
cut on $\Delta E$ and $E$, which is shown in Fig.~\ref{fig:banana2} (red 
dashed contour). The selection is not sensitive enough to resolve between 
the \Be\  emitted at the ground state in the
p($^{7}$Li,$^{7}$Be)n reaction and $^{7}$Be* in the first excited state coming from the 
p($^{7}$Li,$^{7}$Be*)n' reaction. Therefore, the neutron energy distribution consisted of two different mono-energetic 
components.

The data sample was further selected by using the time-of-flight (ToF) of the \TPC\ with respect to the 
$\Delta E$/$E$ telescope, namely by keeping the events in which the 
delay between the telescope and the \TPC\ (see inset in Fig.~\ref{fig:banana2}) is consistent with the 
flight time of the neutrons.
The coincidence window in ToF was set to be S1-dependent, in order 
to ensure a S1-independent selection efficiency. 
The boundaries of the coincidence window were defined as the 1\% 
and 99\% quantiles in each S1 slice of 10 PE, after the subtraction of 
the constant background due to random neutrons and $\gamma$-rays. 
The random background contributes to about \SI{1}{\percent} of the events 
in the coincidence windows. 

\begin{figure}
    \centering
    \includegraphics[width=1.0\columnwidth]{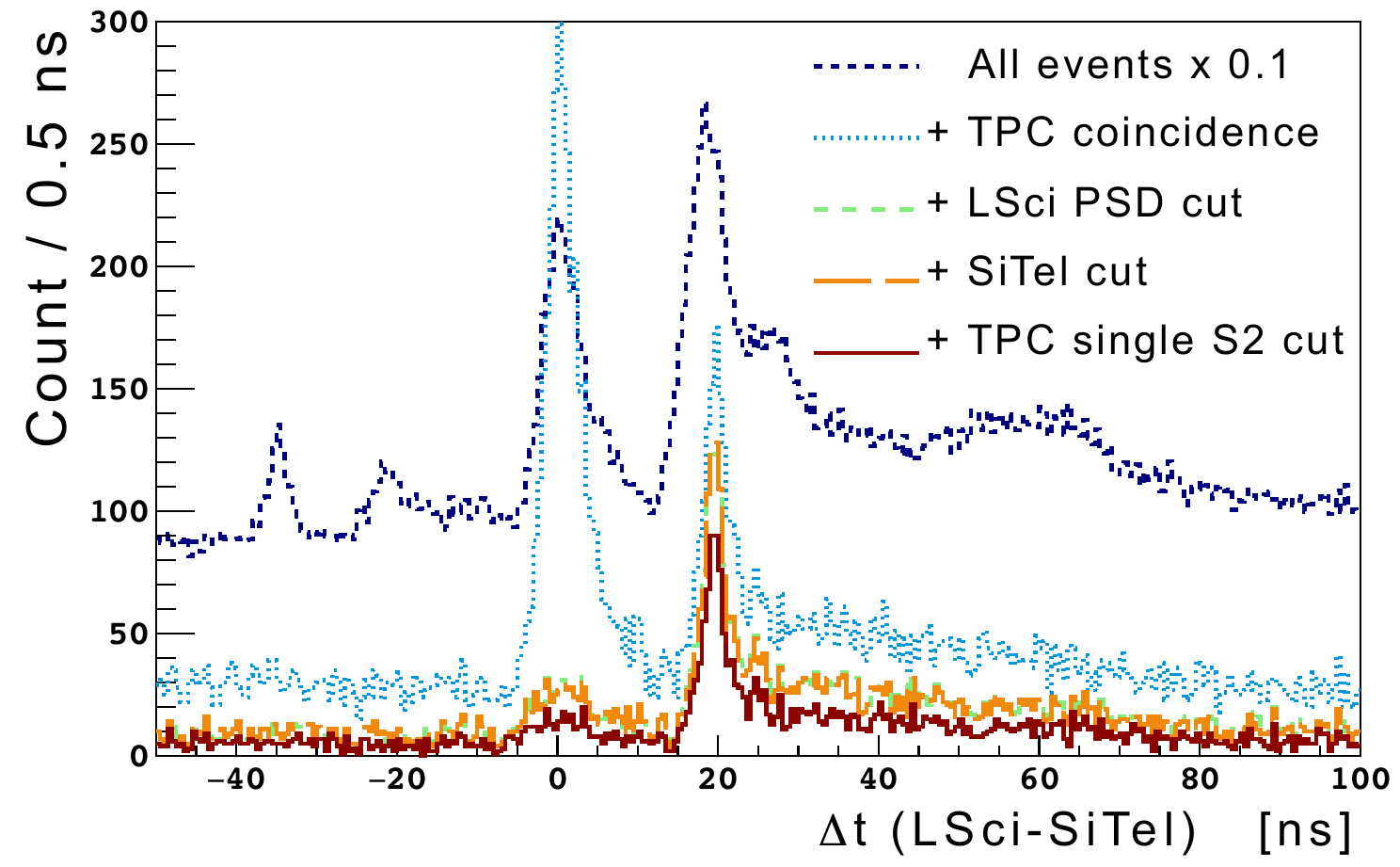}
    \caption{Timing spectra of $\Delta t(\mathrm{LSci-SiTel})$. All channels of the neutron spectrometer are aligned with the $\gamma$ peak at $t=0$. Cuts are applied accumulatively starting from the raw distribution (dark blue dashed histogram). }
    \label{fig:LAr:ReD_Cut_Eff}
\end{figure}

%
The coincidence windows for the delay $\Delta t(\mathrm{LSci}-\mathrm{SiTel})$ 
between the LSci and the telescope in triple-coincidence events were set 
with very stringent cuts, so 
to guarantee the selection of pure single-scattering neutron interactions.  
The timing of the individual LScis was calibrated by using as a reference
the $\gamma$-rays produced in the \TPC\ by inelastic interactions (n,n'$\gamma$) and
then detected in the LScis: all $\gamma$ peaks were aligned 
to $\Delta t(\mathrm{LSci}-\mathrm{SiTel}) = 0$, as displayed in 
Fig.~\ref{fig:LAr:ReD_Cut_Eff}, where the effect of used cuts applied sequentially is shown. 
The single-scattered neutron events of 
interest form the peak around \SI{20}{ns}. The low-statistics peak at about 
\SI{25}{ns} comes from the lower-energy neutrons produced in the 
p($^{7}$Li,$^{7}$Be*)n' interactions, while the tails at longer times are 
mostly due to multi-scattered neutron background. Monte Carlo simulations
indicate that the hump around \SI{60}{\nano\second} is originated by
the neutrons associated with the ``high energy'' \Be, which reach the \TPC\ after 
scattering on the floor or other passive structure. The peaks around 
\SI{-35}{\nano\second} and \SI{-20}{\nano\second} are 
$\gamma$-rays emitted by p($^{7}$Li,$^{7}$Li*)p inelastic scattering. 
Gaussian fits to the peak around \SI{20}{ns} determined the position and 
width of the window, individually for each scintillator. As mentioned 
in Sect.\,\ref{sec:spectrometer}, the LSci channels 
which selected \NR\ events at $\theta_r=\SI{20}{\degree}$ and 
$\SI{40}{\degree}$ were each made from the analogue sum of the signals 
of two different detectors. Since the cable lengths for the two detectors 
at $\SI{20}{\degree}$ were not properly matched, this introduced a split 
in the timing: the  $\Delta t(\mathrm{LSci}-\mathrm{SiTel})$ distribution 
for the channel at \SI{20}{\degree} was hence fitted with a double Gaussian. 
The coincidence windows were defined according to the position $\mu$ and width $\sigma$
of the peaks from the Gaussian fits, as summarized in Table~\ref{tab:LSci_Timing} and they are used to select the triple coincidence events. The
coincidence windows were further extended by \SI{5}{ns} in order to include the slower
neutrons from p($^{7}$Li,$^{7}$Be*)n'. Side-bands were also defined to estimate the 
random coincidence rate in each channel, see Tab.~\ref{tab:LSci_Timing}. 
 
\begin{table*}
    \centering
     \caption[Definition of the coincidence windows in the ToF $\Delta t(\mathrm{LSci}-\mathrm{SiTel})$ for each LSci channel.]{Coincidence and side-band windows in the ToF $\Delta t(\mathrm{LSci}-\mathrm{SiTel})$ for each LSci channel. $d$ is the total width of the coincidence window, $d=6 \sigma + 5$~ns.}
    \label{tab:LSci_Timing}
\begin{tabular}{c|c|c|c|c|c}
\toprule
Angle $\theta_r$ of the \TPC\ \NR  & 90\textdegree $l$ & 40\textdegree & 0\textdegree & 90\textdegree $r$ & 20\textdegree \\
\hline
    Neutron peak  $\mu$ [\si{ns}]   & 19.75  & 19.44   & 19.51   & 20.09   & \makecell{$\mu_1=17.18$, $\mu_2=20.44$} \\
    Timing resolution  $\sigma$ [\si{ns}] & 1.12 & 1.12  & 1.50    & 1.25    & 1.17      \\
    \hline
    Coincidence window        & \multicolumn{4}{c|}{$[\mu-3\sigma,\mu+3\sigma+\SI{5}{ns}]$} & \makecell{$[\mu_1-3\sigma$, $\mu_2+3\sigma+\SI{5}{ns}]$} \\ 
    \hline
    Side-band window          & \multicolumn{5}{c}{$[-\SI{20}{ns}-20d,-\SI{20}{ns}]\cup[\SI{70}{ns},\SI{70}{ns}+20d]$}\\
\bottomrule
\end{tabular}
\end{table*}
The triple coincidence events eventually considered for the statistical analysis of 
Sect.~\ref{sec:StatisticalAnalysis} are those which pass the sequence of cuts displayed 
in Fig.~\ref{fig:LAr:ReD_Cut_Eff} and the additional selection in the $\Delta t(\mathrm{LSci}-\mathrm{SiTel})$ 
ToF from Table~\ref{tab:LSci_Timing}.


\section{Statistical analysis} \label{sec:StatisticalAnalysis}
\begin{figure}
    \centering
    \includegraphics[width=1.0\columnwidth]{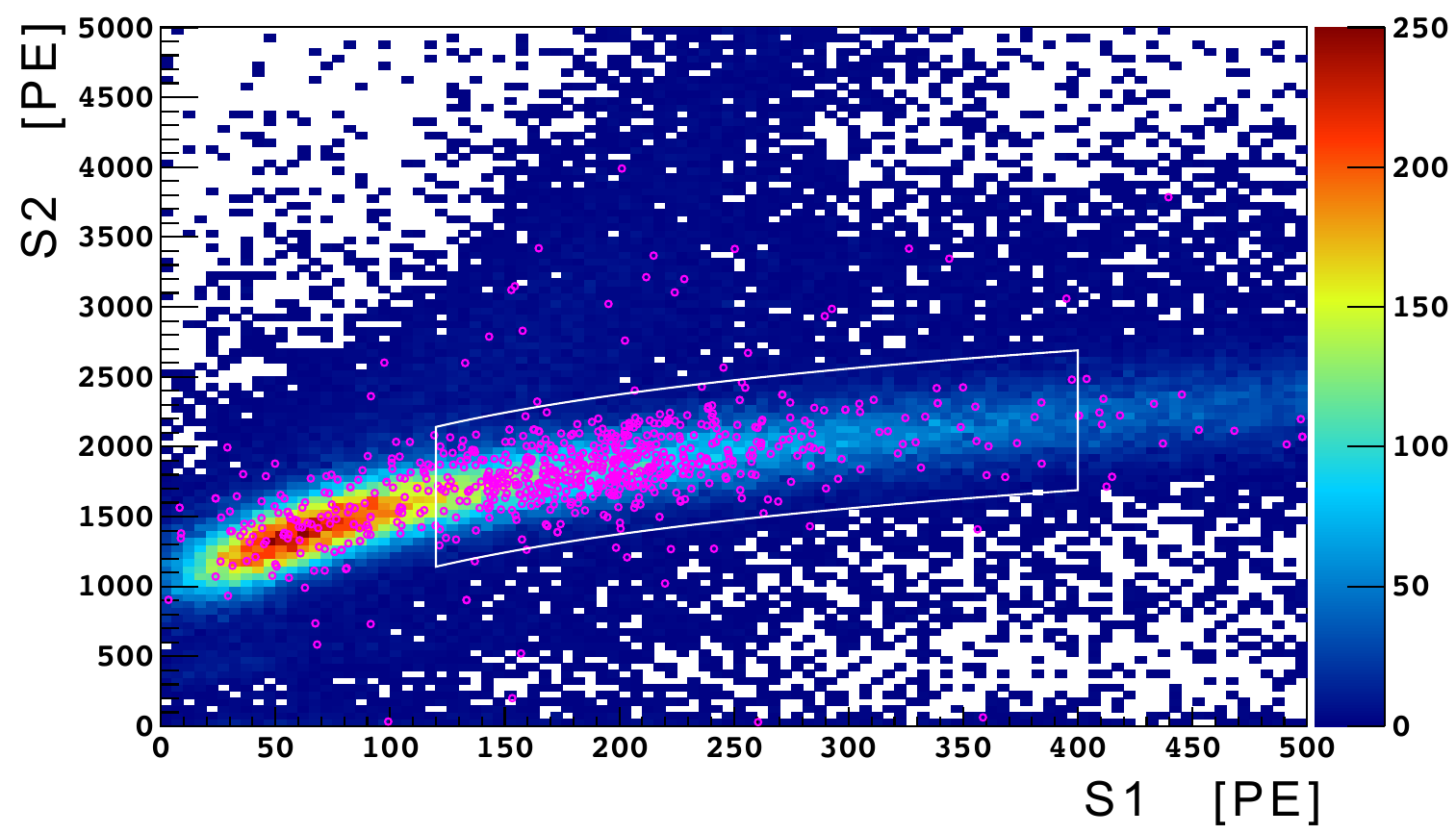}
    \caption{S2 vs. S1 distribution of \NR\ events in the \TPC. The color-coded histogram includes the event in 
    double coincidence (\TPC\ and telescope), namely at all angles $\theta_r$ with respect to the electric field. The 
    pink circles are the events in triple coincidence (\TPC, telescope and spectrometer). All corrections and cuts are applied. The white contour is the fit 
range in the (S1,S2) plane used for the statistical analysis described in 
Sect.~\ref{sec:StatisticalAnalysis}.}
    \label{fig:LAr:ReD_Cut_Coinci}
\end{figure}

The S2 vs. S1 distribution of the \NR\ events in the \TPC\ which pass the 
selection procedure of  Sect.~\ref{sec:EventSelection} is displayed in Figure~\ref{fig:LAr:ReD_Cut_Coinci}:  
the pink dots represent the events selected requiring the triple coincidence (\TPC, Si telescope and neutron 
spectrometer); 
the colour-coded distribution includes the events in double coincidence (\TPC\ and telescope). 
The triple coincidence sample contains about 650 \NR\ events with S1 above \SI{20}{PE}, which were collected 
during 10.7~live days of beam run. The double coincidence events constitute a large sample of about 
70000 \TPC\ \NR\ events in all directions: they were hence
used as a calibration data set to constrain the nuisance model parameters in the global fit below. Since the triple coincidence events are a 
small fraction of the double coincidences, the large sample of double coincidence events was also used 
as the template for random coincidence background. 


The data samples were statistically
analysed in order to evaluate the best estimate of the directionality
parameter $\delta R = R-1$, which measures how much the shape of the
initial ionization charge cloud differs from a sphere. As the number
of events is relatively modest, an unbinned profile likelihood was applied. The
global likelihood $\mathcal{L}$ is written as a product of three likelihood terms:
\begin{eqnarray}\label{eq:LAr:ReD_likelihood_total}
    \mathcal{L}(\vect{X}\,|\,\delta R, \vect{\nu}) & = & \prod_{i=1}^{5} \mathcal{L}_{i}(\vect{X}_i\,|\,\delta R,\theta_{r}^{(i)},\vect{\nu}) \nonumber\\
    & &  \times \mathcal{L}_\mathrm{cali}(\vect{X}_\mathrm{cali}\,|\,\vect{\nu})
    \times \mathcal{L}_\mathrm{constraint}(\vect{\nu}),
\end{eqnarray}
where the product over $i$ refers to the five
samples taken at the five angles $\theta_r^{(i)} = {0^\circ, 20^\circ, 40^\circ, 90^\circ l, 90^\circ r}$ of Tab.~\ref{tab:LSci_Timing}, each containing the observed
array of events $\vect{X}_i = (\mathrm{S1},\mathrm{S2})$; $\delta R$ is the parameter of interest (POI); $\vect{\nu}$ is the array of nuisance parameters; $\vect{X}_\mathrm{cali}$ is the array of calibration data set.
The POI is constrained in this work to $\delta R \ge 0$, as negative values of $\delta R$ are not
physically allowed by the recombination model~\cite{Cataudella:2017kcf}. 
The three  likelihood terms of Eq.~\ref{eq:LAr:ReD_likelihood_total} are described in detail below.

$\mathcal{L}_i$ is the extended likelihood of each sample of \NR\ events at the recoil angle $\theta_{r}^{(i)}$:
\begin{equation}
    \mathcal{L}_i = \mathrm{Poisson}(n_i|\hat{n_i})\prod_{X_j\in\vect{X}_i} \mathcal{P}_i(\mathrm{S1}_j,\mathrm{S2}_j; \delta R, \theta_{r}^{(i)}, \vect{\nu}) 
\end{equation}
where $n_i$ and $\hat{n_i}$ are the size of $\vect{X}_i$ and its mean, respectively, and $\mathcal{P}_i$ is the 
joint probability density function (PDF) of the events (S1,S2). 
The PDF is made as the combination of three components, one for
signal and two from backgrounds:
\begin{eqnarray} \label{eq:eachpdf}
    \mathcal{P}_i(S1,S2)&=& (1-\lambda_{1i})(1-\lambda_2)F_\mathrm{sig}(E_r) \nonumber\\
    & &  \otimes P(\mathrm{S1},\mathrm{S2}; \delta R, \theta_{r}^{(i)}, \vect{\nu}, E_r) \nonumber\\             
    & & + [\lambda_{1i} F_\mathrm{bkg1}(E_r) + (1-\lambda_{1i})\lambda_2 F_\mathrm{bkg2}(E_r) ]\nonumber \\
    & & \otimes P(\mathrm{S1},\mathrm{S2}; \delta R, \bar{\theta}_r, \vect{\nu}, E_r).
\end{eqnarray}
The first component is the energy spectrum for the signal $F_\mathrm{sig}(E_r)$, which
depends on the recoil energy $E_r$, convolved with the response function $P$ of the
\TPC\ to mono-energetic \NR\ events, as defined in Eq.~\ref{eq:TPCResponse}. The parameters $\lambda_{1i}$ are the fractions of random coincidences within 
each data sample: they were estimated from the data, using the counting rate
in the side-band in ToF and are listed in Tab.~\ref{tab:LAr:lambda_1}.
Similarly, $\lambda_2$ is the scaling factor for multi-scattering background, namely
the fraction of those events with respect to all \NR\ events in the coincidence
window.  The other
two components are the energy distributions of the backgrounds due to random
coincidences, $F_\mathrm{bkg1}(E_r)$, and to multiple neutron scattering, 
$F_\mathrm{bkg2}(E_r)$. They are also convolved with the response function $P$ of
the \TPC. 

\begin{figure}
    \centering
    \includegraphics[width=\linewidth]{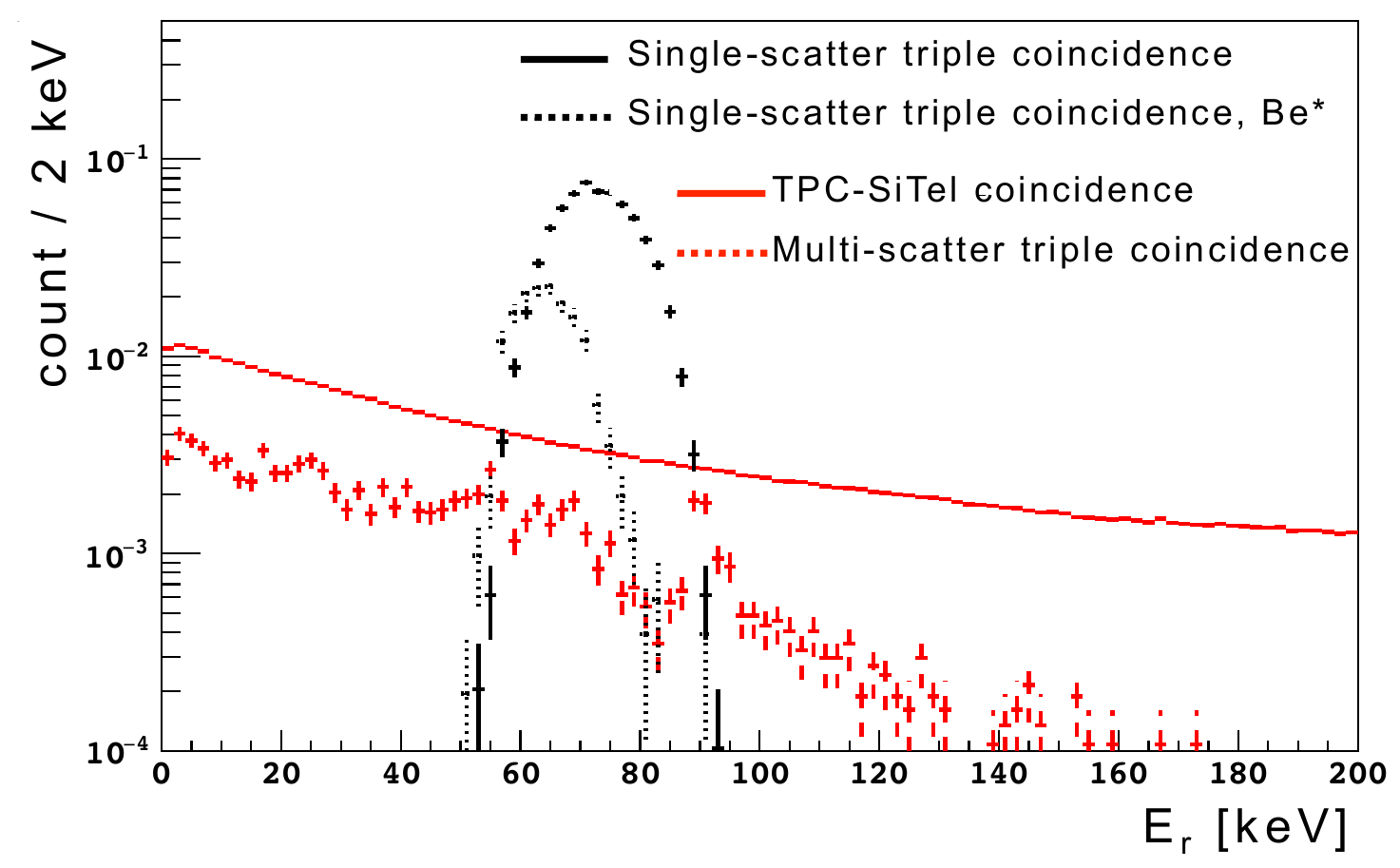}
    \caption{Signal and background spectra from \texttt{g4ds} Monte Carlo simulation. Solid and dotted black lines show the
    distribution of the \NR\ energy $E_r$ for signal events associated with \Be\ and $^7$Be$^*$ neutrons,
    respectively. Solid red line shows the random coincidence background spectrum $F_\mathrm{bkg1}$ and the dashed red
    line the spectrum $F_\mathrm{bkg2}$ from multi-scattered triple coincidence events. The error bars represent the
    Monte Carlo statistical uncertainties.}
    \label{fig:simspectra}
\end{figure}

In order to speed-up the computation of the response function $P$, the
Poisson and binomial distributions are approximated by Gaussian distributions, such
that the convolutions over $N_\mathrm{ph}$ and $N_{e^-}$ in Eq.~\ref{eq:TPCResponse} can
be evaluated analytically.
As the angular distribution for background events is approximately random, the
$\theta_r$ dependence of $f(\theta_r, R)$ is averaged out by using the equivalent
angle $\bar{\theta}_r$ calculated analytically for an isotropic distribution and
the functional dependence on the angle is approximated as
$\langle f(\theta_r, R)\rangle \sim f(\bar{\theta}_r, R)$.
The factor $\lambda_2$ and the three energy spectra ($F_\mathrm{sig}$,
$F_\mathrm{bkg1}$, and $F_\mathrm{bkg2}$) were evaluated by means of a dedicated 
Monte Carlo simulation using the \texttt{Geant4}-based framework \texttt{g4ds}~\cite{Agostinelli:2003fg,Allison:2006cd,Allison:2016lfl,Agnes:2017grb}.
The events from the simulations
underwent the same sequence of selection cuts used for the
real data. The energy distributions derived by the Monte Carlo are displayed
in Fig.~\ref{fig:simspectra}. The three energy distributions were then analytically
parametrized in order to optimize the calculation of the CPU-intensive
PDF $\mathcal{P}_i$. $F_\mathrm{sig}$ consists of two Gaussian peaks corresponding
to the \NR\ induced by neutrons from p($^{7}$Li,$^{7}$Be)n and 
p($^{7}$Li,$^{7}$Be*)n'.
$F_\mathrm{bkg1}$ and $F_\mathrm{bkg2}$ were approximated by a double-exponential and a
single exponential, respectively, whose parameters were calculated by fits to
the Monte Carlo distributions.
%

\begin{table}
    \centering
    \caption{Fraction of random coincidence events, $\lambda_{1i}$, in the range
    S1$\,\in[120,400]\si{PE}$ and S2$\,\in[800,2800]\si{PE}$ in the five samples of
    triple-coincidence events at different $\theta_r$. Uncertainty is about 2\% for all 
    samples.}
    \label{tab:LAr:lambda_1}
    \begin{tabular}{c|c|c|c|c}
    \toprule
    \thead{0\textdegree} & \thead{20\textdegree} & \thead{40\textdegree} &  \thead{90\textdegree $l$} & \thead{90\textdegree $r$} \\
    \hline
    0.045 & 0.048 & 0.047 & 0.026 & 0.041 \\
        \bottomrule
    \end{tabular}
\end{table}

The factor $\mathcal{L}_\mathrm{cali}$ of the global likelihood of Eq.~\ref{eq:LAr:ReD_likelihood_total} is the constraint
term on the nuisance parameters and it depends on the events $\vect{X}_\mathrm{cali}$ in the calibration set (i.e. colour-coded histogram in Fig.~\ref{fig:LAr:ReD_Cut_Coinci}) 
While the energy spectrum of the calibration events is 
a broad and featureless distribution, the joint distribution of the \NR\ band in the (S1,S2) plane can set a
strong constraint on the nuisance parameters. Since the fraction of signal events in the calibration sample is negligible,
the energy distribution is well approximated by the random background $F_\mathrm{bkg1}$. The calibration term is hence
written as:
\begin{equation}
\label{eq:L_cal}
    \mathcal{L}_\mathrm{cali} = \prod_{X_j\in\vect{X}_\mathrm{cali}} P(\mathrm{S1}_j,\mathrm{S2}_j; \delta R, \bar{\theta}_r, \vect{\nu}, E_r)\otimes F_\mathrm{bkg1}(E_r).
\end{equation}

In order to avoid any analysis bias, $\delta R$ should be decoupled from the nuisance parameters as much as possible. The
explicit occurrence of the POI $\delta R$ in Eq.~\ref{eq:L_cal} is due to the fact that the parameter $\xi_m$ in
the modified Thomas-Imel model in Eq.~\ref{eq:rec_direction} is dependent on $\delta R$ because of the track length. To
remove such undesirable degeneracy, the angular dependence term and the Thomas-Imel parameter were re-defined as 
\begin{equation}
f'(\theta_r,R) = f(\theta_r,R)/f(\bar{\theta}_r,R)
\end{equation}
and
\begin{equation}
\xi_m' = \xi_m/f(\bar{\theta}_r,R),
\end{equation}
respectively. In this way the angle-averaged position of the \NR\ band in calibration data does not depend on
$\delta R$ and the POI $\delta R$ is left as a pure representation of directionality. Furthermore, the degenerate
nuisance parameters were re-cast into a unique nuisance parameter $A=\xi_m'/(\mathcal{E}_d \cdot \langle N_\mathrm{i} \rangle)$, which represents the
recombination probability of one electron-ion pair. 

The last factor of the global likelihood,
$\mathcal{L}_\mathrm{contraint}(\vect{\nu})$, is the pull term for the nuisance parameters which were known 
by prior independent measurements. Those parameters are constrained by Gaussian terms
\begin{equation}
    \mathcal{L}_\mathrm{constraint}(\vect{\nu}) = \prod_i\frac{1}{\sqrt{2\pi}\sigma_{\nu_i}}\exp{-\frac{(\nu_i-\nu_i^0)^2}{2\sigma_{\nu_i}^2}}
\end{equation}
based on the previously-measured values $\nu_i^0$ of the parameters $\nu_i$ and on their corresponding uncertainties.


\begin{table}
    \centering
    \caption[Parameter values in the signal model in ReD directionality analysis.]{List of the parameters used in the model. $\delta R$ is the parameter of interest, while all others are nuisance parameters, constrained by the calibration data and/or by a Gaussian pull term. The error bars are the standard deviation which is taken in the Gaussian pull terms. The parameters reported without uncertainties are fixed. The gains $g_1$ and $g_2$ come from the previous \TPC\ performance study~\cite{Agnes:2021zyq}. The S1 resolution of the \TPC\ of Eq.~\ref{eq:TPCResponse} is parametrized as $\sigma_{\mathrm{S1}}^2 = \mathrm{S1}/[\si{PE}] + {\sigma^*_{\mathrm{S1}}}^{2} $, namely by the combination of the statistical term and of an extra contribution. The same is done for the S2 resolution.}
    \label{tab:parameters}
    \begin{tabular}{c|c|c}
    \toprule
         & \thead{Constraint}   & \thead{Comment} \\
        \hline
        $\delta R$  & -        & Parameter of interest \\
        \hline
        $A$    &   \makecell{$0.04\pm0.01$\\ $[\si{1/e^-}]$ }    
        & $A = e/[2\pi\epsilon_r\epsilon_0\mathcal{E}_d\sigma^2 \mu_- f(\bar{\theta}_r,R)]$\\
        \hline
               $k_e$       & $2.8$          
        & Electronic quenching coefficient~\cite{Mei:2008ca}\\
\hline
$W_{ph}$         & \makecell{$19.5$\\ $[\si{eV}]$}       & \makecell{Energy  for scintillation \\photon production~\cite{Doke:2002oab}}\\
\hline
        $N_\mathrm{ex}/N_\mathrm{i}$ & $0.2\sim2$  
        & \makecell{Excitation to ionization ratio.\\ Energy dependence as in \cite{Cao:2015ks}} \\
        \hline
        $g_1$       & \makecell{$0.196\pm0.020$\\ $[\si{PE/ph}]$}      & S1 signal yield \\
        \hline
        $g_2$       & \makecell{$20.5\pm2.5$\\ $[\si{PE/e^-}]$}       & S2 signal yield \\
        \hline
        $\sigma^*_{\mathrm{S1}}/\mathrm{S1}$ &  $0.003\pm0.05$    
        & \makecell{S1 detector resolution\\ in addition to $\sqrt{\mathrm{S1}}$} \\
        \hline
        $\sigma^*_{\mathrm{S2}}/\mathrm{S2}$ &  $0.001\pm0.05$    
        & \makecell{S2 detector resolution\\ in addition to $\sqrt{\mathrm{S2}}$} \\
        \hline
        $\lambda_1$ &  Table~\ref{tab:LAr:lambda_1}      & Fraction of random coincidence \\
        \hline
        $\lambda_2$ &   0.16        & \makecell{Ratio of multi-scattering to \\all NR in coincidence windows}\\
        \bottomrule
    \end{tabular}
\end{table}

As a summary, the parameters and their reference values are summarized in Tab.~\ref{tab:parameters}.
The recombination probability $A$ depends on $\sigma$, the size of the ionization cluster of 
Eq.~\ref{eq:clouddist}, which is dominated by
the electron diffusion during thermalization. Due to their high mobility and long thermalization time,
electrons diffuse for a few \textmu m in LAr \cite{Wojcik:2003ja,mozumder1995free}. It is found
that $A=0.04/e^-$, which corresponds to $\sigma=\SI{1.8}{\micro\meter}$, was an appropriate initialization parameter
for the likelihood fit. The ratio $N_\mathrm{ex}/N_\mathrm{i}$ was treated as a function of recoil energy, according to
the indications by SCENE~\cite{Cao:2015ks}. The \TPC\ gains $g_1$ and $g_2$ were estimated according to the \TPC\
characterization in~\cite{Agnes:2021zyq}, and were treated as nuisance parameters in order to accommodate for
possible variations in the \TPC\ performance. The parameters $W_{ph}$, $k_e$, $\lambda_1$ and $\lambda_2$ were fixed
in order to limit the degeneracies in the fit: their effect on the POI is minor and is accounted
below as a systematic uncertainty. 

Finally, experimental data of Fig.~\ref{fig:LAr:ReD_Cut_Coinci} (calibration 
and five triple-coincidence samples) were fitted against the model of
Eq.~\ref{eq:LAr:ReD_likelihood_total}. In order to make the fit stable the 
fit region in the (S1,S2) plane was selected in order to include only the 
\NR\ band, with S1$\,\in[120,400]\si{PE}$, as represented by the white contour in 
Fig.~\ref{fig:LAr:ReD_Cut_Coinci}. The S1 range corresponds to NR energies between
approximately 35 and 150~keV, and hence comfortably includes the expected NR signal at
$\sim 72$~keV. The low-S1 edge $S1 > 120~\si{PE}$ was set in order to avoid any inefficiencies in
the event reconstruction and selection. The center of the \NR\ band
was empirically parametrized with the function S2$\,/[\si{PE}] = 455\ln(\mathrm{S1}/[\si{PE}])-535$
and the cut was set as $\pm500~\si{PE}$ in S2. The fit region globally contains 
529 triple coincidence and 42340 calibration events.

\begin{figure*}
    \centering
    \includegraphics[width=1.0\linewidth] {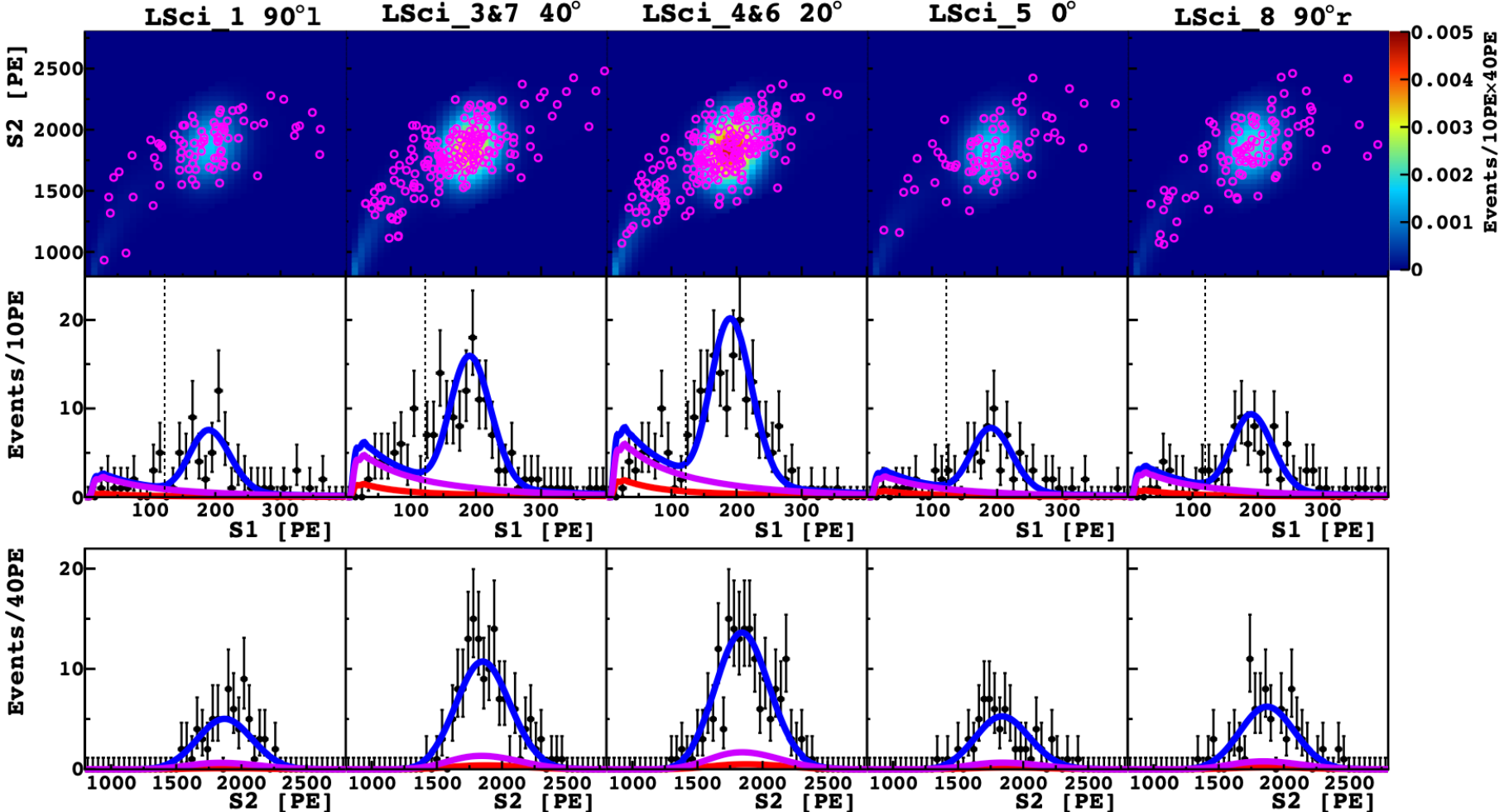} 
    \caption[The best-fit of the ReD directionality likelihood analysis]{Experimental data superimposed with the 
best-fit model. The fit is performed for $S1 \in [120,400]\si{PE}$ and within the white contour of 
Fig.~\ref{fig:LAr:ReD_Cut_Coinci}. Upper row: S2 vs. S1 distribution for the
calibration data set (color-coded histogram) and for 
the triple-coincidence data sets at different angles $\theta_r$ (pink circles).  Middle row: projection on S1 for the 
triple-coincidence samples. The dashed vertical lines at $120 \si{PE}$ mark the left edge of the fit range.
Bottom row: projection on S2 for the 
triple-coincidence sample in the range of $S1\in[120,400]\si{PE}$. The blue, red and pink curves are the total spectrum, 
the random coincidence background $F_\mathrm{bkg1}$, and multi-scattering background $F_\mathrm{bkg2}$, respectively. }
    \label{fig:ReD_Fit_result}
\end{figure*}

\begin{table}
    \centering
    \caption{Best-fit of the parameters and correlation coefficients between the nuisance parameters 
and the POI $\delta R$.}
    \label{tab:Fit_result}
    \begin{tabular}{c|c|c}
    \toprule
 \thead{Parameter} & \thead{Value}  & \thead{Correlation with $\delta R$} \\
 \hline
 $\delta R$ & $0.037\pm0.027$ & -  \\
 $A\,[\si{1/e^-}]$ & $(4.01\pm0.06)\times10^{-2}$ &-0.014  \\
 $g_1\,[\si{PE/ph}]$ & $0.204\pm0.002$ & 0.013  \\
 $g_2\,[\si{PE/e^-}]$& $20.1\pm0.2$ & -0.009  \\
 $\sigma^*_{\mathrm{S1}}$/S1  & $0.017\pm0.003$ & -0.012  \\
 $\sigma^*_{\mathrm{S2}}$/S2  & $0.0002\pm0.0060$ & 0.026  \\
 \bottomrule
    \end{tabular}
\end{table}

The fit result is shown in Fig.~\ref{fig:ReD_Fit_result} and reported in Tab.~\ref{tab:Fit_result}.
The positions of the signal peak in both S1 and S2 (middle and bottom rows of Fig.~\ref{fig:ReD_Fit_result}) are mutually 
consistent among the five samples at different $\theta_r$. The best-fit for the POI is 
$\delta R = 0.037\pm0.027$, which is less than $2\sigma$ away from a null result; 
the uncertainty on $\delta R$ is largely driven by statistics. 
The upper limit of 
$\delta R$ is calculated by a toy Monte Carlo approach, in order to guarantee the correct coverage: it results to be
$\delta R < 0.072$ at 90\% CL. 
The best-fits of the nuisance parameters are in good agreement with the central values of their 
estimates used for the constraints.
In particular, the smallness of the best-fit for the parameters 
$\sigma^*_{\mathrm{S1}}$/S1 and $\sigma^*_{\mathrm{S2}}$/S2, which are the extra (non-statistical) contributions to the 
experimental resolution in S1 and S2, demonstrates that the spatial inhomogeneities of the detector response 
were properly corrected. Furthermore, the proper convergence and the absence of a significant bias for all
fit parameters, notably including the POI $\delta R$, were checked by running a dedicated set of toy Monte Carlo
simulations.

The uncertainties on $\delta R$ related to the nuisance parameters are automatically accounted in 
the fit. All other systematic uncertainties on $\delta R$, e.g. those related to the values of $W_{ph}$, $k_e$,
$\lambda_1$ and $\lambda_2$, to the spectral shapes $F_\mathrm{sig}$, $F_\mathrm{bkg1}$ and $F_\mathrm{bkg2}$, and to
the approximation of  $\bar{\theta}_r$ from isotropic distribution, are globally evaluated to be an order
of magnitude smaller than the statistical term and are hence neglected in this work.

\section{Discussion} \label{sec:discussion}
The results of this work suggest that the charge recombination in \NR s in the energy 
range of interest for WIMP DM searches has a limited directional dependence.
A possible explanation is that the directional effect is washed out in 
the isotropic thermalization process of the electrons: the range of 
\SI{70}{\kilo\electronvolt} argon ions in LAr, 
\SI{0.18}{\micro\meter} \cite{srim}, is much shorter than the 
electron thermalization radius 
$r_t \sim $\SI{2.5}{\micro\meter}~\cite{Wojcik:2003ja,Wojcik:2016gy}. 
If all electrons were confined within the Onsager radius, the 
recombination probability $A$ would be $8/e^-$, namely, 
two orders of magnitude higher than measured in this work. This
indicates that the extension of the thermalized electron cloud is much bigger
than the Onsager radius, thus weakening the initial directional
effect. Other non-local processes at the length scale of a few \textmu m 
can also contribute to the size of the electron cloud, including the emission of
Auger electrons and fluorescence X-rays from excited Ar atoms~\cite{estar,xcom}.

The strongest constraint on $\delta R$ from the fit comes from the position 
of the S2 peak, since $g_2\gg g_1$. In fact, the SCENE hint for directional 
sensitivity was primarily given by the 7\% variation in S1 for \NRs\ parallel and perpendicular 
to \edrift: no variation of S2 vs. direction was 
observed in SCENE. 
While the SCENE data were never analyzed according to the directional
model of \cite{Cataudella:2017kcf}, an asymmetry $\delta R \approx 2$ would be required
to generate a 7\%-effect on S1. However, given the anti-correlations of Eqs.~\ref{eq:ne}
and \ref{eq:nph}, such a large $\delta R$ would produce a much more significant variation in S2
($\sim 80\%$ between parallel and perpendicular directions), which is not observed
in SCENE. The lack 
of a variation in the S2 signal, which is further confirmed in this work, 
rules out the directional modulation in charge recombination as the 
explanation of the effect and sets an upper limit on $\delta R$. 
Furthermore, the \ReD\ data, with an improved signal yield and resolution 
in S1, do not confirm the variation in S1 at different directions 
which was reported by SCENE. As for S2, no statistically-significant 
variation was found for S1.

The LAr signal model adopted in this work has two major upgrades
comparing to the models commonly used in the literature. 
The first modification is about charge recombination, by the introduction 
of the directional term of Eq.~\ref{eq:rec_direction} from 
Ref.~\cite{Cataudella:2017kcf}. The second modification is the use of an 
energy-dependent ratio \NexNi, which allows for 
a better fit of the \NR\ band shape and improves significantly the 
performance of the likelihood fit.
If \NexNi\ is kept constant to the value 1, which is commonly-adopted for NRs~\cite{Doke:2002oab,Hitachi:2021hac},
the fit still returns a value of $\delta R$ compatible with zero, but the model fails to reproduce the shape 
of the S2 vs. S1 band and the S2 distribution for \NR s; furthermore, the best-fit for $g_2$ in this case 
is $29.9\pm0.1$~\si{PE/e^-}, which is in tension with the prior measurement of Table~\ref{tab:parameters}.
%
%
While the SCENE data also support the energy dependence of \NexNi, the physical motivation of it requires further 
study. One possibility is that this is the
apparent effect of energy dependences in the nuclear quenching, 
electron quenching and recombination processes, which are unaccounted by
the Lindhard, Mei and Thomas-Imel models used in this work. Specifically:
the Thomas-Fermi screening function used in the Lindhard model is known to
have a bias in the $O(10)\si{keV}$ range~\cite{Bezrukov:2010qa,Sigmund2004}; the Mei model simplifies the
average electronic stopping power by taking the value at the initial electron
kinetic energy; the charge recombination model does not consider the dependence
on the charge cloud size on the recoil energy. All these energy-dependent
factors are not accounted in the models and they could eventually show up
as an effective energy dependence in \NexNi.
It has anyhow a small effect on the systematic uncertainty on $\delta R$, 
due to the weak correlation reported in Tab.~\ref{tab:Fit_result}. 

Taking the central value $\delta R=0.037$ from the fit, the amplitude of the S2 daily 
modulation curve, which results by folding the directional dependence 
S2$(\theta)$ of this work into the WIMP recoil direction distribution 
from Ref.~\cite{Cadeddu:2017ebu}, is only \SI{0.5}{\percent} peak-to-peak. 
Given the very large number of candidate events required to 
statistically identify it, this effect could hardly be of experimental 
use in a future $O(100)$~ton argon dual-phase \TPC\ DM detector. 


\section{Conclusions} \label{sec:conclusions}
The Recoil Directionality (\ReD) experiment was designed within the 
GADMC to explore the possible directional sensitivity of an Ar dual-phase \TPC\ to 
nuclear recoils in the energy range of interest for WIMP DM searches.
The \ReD\ \TPC\ was 
irradiated with neutrons of known energy and direction at the INFN, 
Laboratori Nazionali 
del Sud, in order to produce Ar recoils of about 70~keV kinetic energy. Nuclear 
recoils traveling in five different directions with respect to the drift field
\edrift\ of the \TPC\ were selected using a neutron spectrometer made by 
liquid scintillation detectors. A statistical analysis based on the 
Cataudella et al. model~\cite{Cataudella:2017kcf} was performed to 
assess the  \TPC\ response for those samples of \NR\ 
events. 

The data from this work do not show any statistically-significant 
dependence of either S1 or S2 on the direction for \NRs\ of $\sim 70$~keV.
The best-fit for the parameter of interest $R$, which measures the aspect ratio 
between the long and short axes of the initial electron cloud, is
$R = (1.037 \pm 0.027)$, or $R < 1.072$ at 90\% CL. The absence of 
significant deviations from the spherical symmetry of the electron cloud
indicates that the electron thermalization process likely plays a 
significant role in weakening any initial direction-induced anisotropy 
of the charge cloud. 

\appendix

\section{A data-driven analysis approach}
The directionality analysis presented in this work depends on 
the specific model by Cataudella et al.~\cite{Cataudella:2017kcf} which 
was adopted to describe the phenomenon. One possibility to release such a 
model depencence and to validate the result is to employ a data-driven 
approach based on Machine Learning (ML) techniques. ML techniques are in 
fact very effective in revealing possible correlations between quantities 
in the study of phenomena for which large data samples are available, 
even if a model for their description is lacking.

Supervised learning algorithms were used to try to highlight the signature for possible directionality effects in the electron-ion recombination in the \ReD\, data~\cite{noemipthesis}. 
Due to the limited size of the triple coincidence event samples, an indirect approach was adopted, which makes use of all \TPC\ calibration events. The data set of the double coincidence events provides a two-order-of-magnitude larger amount of data for the training of the model, and this is a desirable condition when working with ML algorithms.

In an ideal \LArTPC, the S2 signal is expected to be related to S1 through some functional form S2$=f$(S1). The basic 
assumption of this strategy is that the function $f$ does not depend on the direction of the Ar recoil, namely that 
the angle $\theta_r$ between the recoil and the drift field does not affect the balance between S1 and S2. 
Deviations from this trend would highlight a possible effect of the recoil directionality.

The model was derived by using the calibration data set, which is made of NR events characterized by a
wide distribution of angles $\theta_r$.  The data set contained about 72000 events and it was randomly split (70:30) in a training 
and testing set, on which the model was trained and tested, respectively. During the training phase, the algorithm built the function $f$ used to predict S2, event-by-event, based on the patterns which are learnt from the training set. 
Each pattern consists of a vector of features: S1 signal [PE], \xy\, position [cm], and \tdrift [\textmu s] as the $z$ coordinate of the event, within the appropriate ranges, and the measured S2 value as a target.
The derived model aims to predict the value of the ionization signal S2, for each of the events, from the knowledge of S1 and of the reconstructed interaction point within the \TPC.

The Extreme Gradient Boosting algorithm (\texttt{xgboost}) was used to derive the model~\cite{XGB}. 
To evaluate the accuracy of the model, the metric of the relative prediction error was adopted. This 
is defined, for the $i$-th pattern (i.e. the $i$-th event in the \TPC), as

\begin{equation}
\epsilon_ {pred}^i =\frac{\mathrm{S2}_{measured}^i - \mathrm{S2}_{predicted}^i}{\mathrm{S2}_{measured}^i}.
\label{par}
\end{equation}

The trend of $\epsilon_{pred}^i$ was investigated for each event, and also against each feature describing the patterns, to verify that there were no regions in the feature space in which the model has a worse response that could introduce any bias in the predictions. 
At the end of the training phase, the model was able to provide a satisfactory prediction of the experimental S2
of the events in the testing set: the relative errors $\epsilon_ {pred}^i$ resulted to be approximately Gaussian-distributed
with mean 0.0043(6) and standard deviation 0.09.
 
\begin{figure}
\centering
\includegraphics[width=0.95\columnwidth]{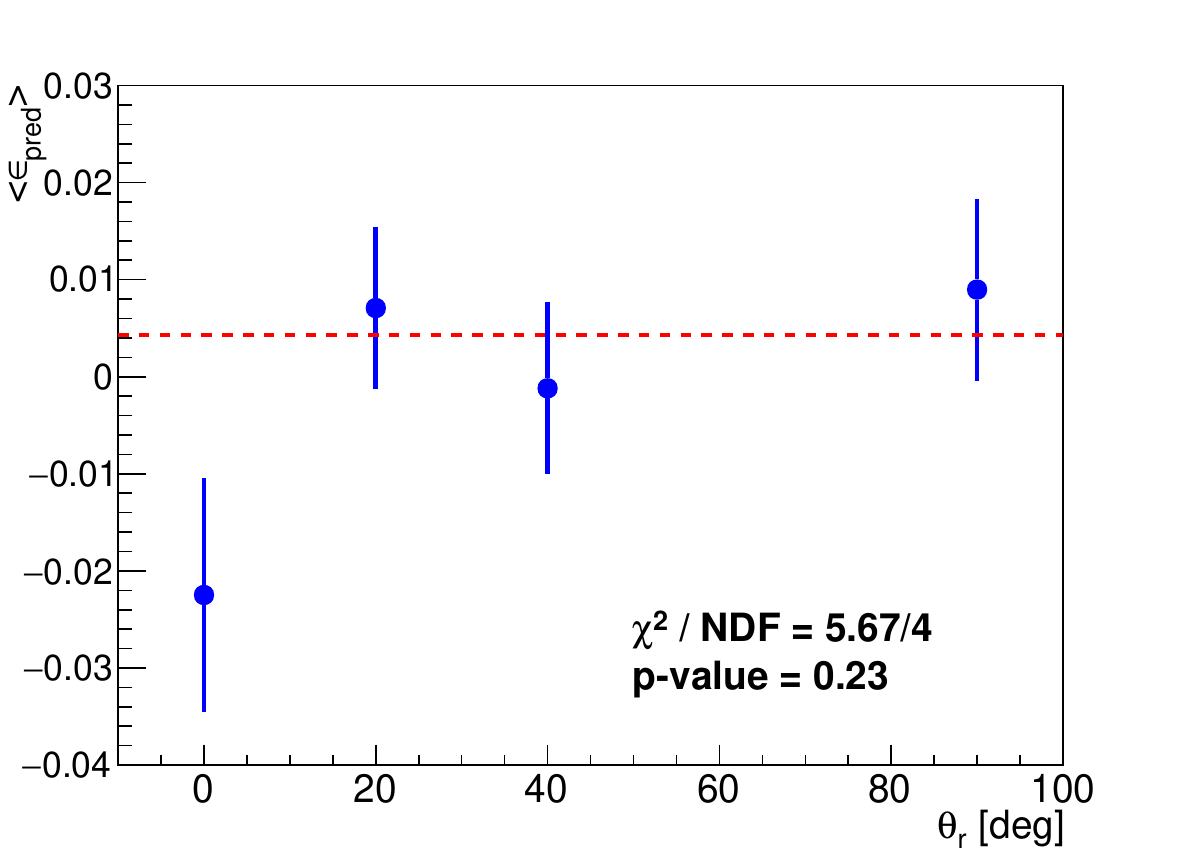} 
\caption{Mean relative prediction error for each $\epsilon _{pred}^i$ distribution obtained by splitting data into the four data sets. The red dashed line marks the level $\epsilon _{pred} = 0.0043$ from the testing set (i.e. null directional effect).}
\label{fig:points}
\end{figure}

Subsequently, the model was used to make predictions on the triple coincidence data set.
For these data, 
the known $\theta_r$ values are used to check for possible directional-dependent deviations of the predicted S2 values compared to those measured experimentally. 
$\epsilon_ {pred}^i$ was initially calculated for each event in the triple coincidence data set and the corresponding values were subdivided into four subsets, according to the angle $\theta_r$ determined by the coincident neutron detection. The mean value of $\epsilon_ {pred}$ in each data set and the corresponding uncertainty are displayed in Fig.~\ref{fig:points} as a function of the recoil direction $\theta_r$.
The point at $\theta_r$ = \SI{0}{\degree} is lower than the others,
as expected in the case of directionality effects, since traces parallel to \edrift\ would result in enhanced S1 signals and reduced S2.
Nevertheless, experimental data are compatible with the null hypothesis of no directionality effect: the $p$-value calculated from the
$\chi^2$ test is 23\%. Therefore, the data-driven analysis carried out using ML techniques on the data collected in the \ReD\ \TPC\ is
compatible with the absence of any directional effect\footnote{A dedicated Monte Carlo based sensitivity study confirmed that, despite the small
size of the triple-coincidence sample, a directional effect as hinted by SCENE (7\% difference in S1 between parallel
and perpendicular recoils) would have been detected by this analysis at 3.2 $\sigma$ level.}, in agreement with the analysis based on
the model by Cataudella et al.~\cite{Cataudella:2017kcf}. 



\begin{acknowledgements}
The Authors express their gratitude to Drs. G.~Cuttone and S.~Gammino, former and current 
Directors of the INFN Laboratori Nazionali del Sud, for the strong and 
constant support to the project. The Authors also thank the entire technical and 
administrative staff of the INFN Laboratori Nazionali del Sud.\\
This report is based upon work supported by the U. S. National Science
Foundation (NSF) (Grants No. PHY-0919363, No. PHY-1004054, No.
PHY-1004072, No. PHY-1242585, No. PHY-1314483, No. PHY- 1314507,
associated collaborative grants, No. PHY-1211308, No. PHY-1314501, No.
PHY-1455351 and No. PHY-1606912, as well as Major Research
Instrumentation Grant No. MRI-1429544), the Italian Istituto Nazionale
di Fisica Nucleare (Grants from Italian Ministero dell’Istruzione,
Università, e Ricerca Progetto Premiale 2013 and Commissione
Scientific Nazionale II), the Natural Sciences and Engineering
Research Council of Canada, SNOLAB, and the Arthur B. McDonald Canadian Astroparticle Physics
Research Institute.\\
We acknowledge the financial support by LabEx UnivEarthS
(ANR-10-LABX-0023 and ANR18-IDEX-0001), Chinese Academy of Sciences
(113111KYSB20210030) and National Natural Science Foundation of China
(12020101004).
This work has been supported by the S\~{a}o Paulo Research Foundation (FAPESP) grants
2018/01534-2 (A.~Sosa), 2017/26238-4 (M.~Ave) and 2021/11489-7. I.~Albuquerque is
partially supported by Conselho Nacional de Desenvolvimento Cient\'{\i}fico e Tecnol\'ogico (CNPq).
The authors were also supported by the Spanish Ministry of Science and
Innovation (MICINN) through the grant PID2019-109374GB-I00, the
``Atraccion de Talento'' grant 2018-T2/TIC-10494,
the Polish NCN (Grant No. UMO-2019/33/B/ST2/02884), the Polish
Ministry of Science and Higher Education (MNiSW, grant number
6811/IA/SP/2018), the International Research Agenda Programme
AstroCeNT (Grant No. MAB/2018/7) funded by the Foundation for Polish
Science from the European Regional Development Fund, the European
Union’s Horizon 2020 research and innovation program under grant
agreement No 952480 (DarkWave), the Science and Technology Facilities
Council, part of the United Kingdom Research and Innovation, and The
Royal Society (United Kingdom), and IN2P3-COPIN consortium (Grant No.
20-152). We also wish to acknowledge the
support from Pacific Northwest National Laboratory, which is operated
by Battelle for the U.S. Department of Energy under Contract No.
DE-AC05-76RL01830.
This research was supported by the Fermi National Accelerator
Laboratory (Fermilab), a U.S. Department of Energy, Office of Science,
HEP User Facility. Fermilab is managed by Fermi Research Alliance, LLC
(FRA), acting under Contract No. DE-AC02-07CH11359.\\
For the purpose of open access, the authors have applied a Creative
Commons Attribution (CC BY) public copyright license to any Author
Accepted Manuscript version arising from this submission.

\end{acknowledgements}

\bibliographystyle{spphys}       
\bibliography{beampaper}   


\newcommand{\notds}{\nolinebreak\footnotemark\nolinebreak}
\renewcommand{\thefootnote}{$*$}
{
\onecolumn
\textbf{The DarkSide-20k Collaboration}
\footnotetext{Not a member of the DarkSide-20k Collaboration}

P.~Agnes\thanksref{AQGSSI}\nolinebreak,
I.~Ahmad\thanksref{AstroCeNT}\nolinebreak,
S.~Albergo\thanksref{CTINFN}{CTUNI}\nolinebreak,
I.~F.~M.~Albuquerque\thanksref{USP}\nolinebreak,
T.~Alexander\thanksref{PNNLaddress}\nolinebreak,
A.~K.~Alton\thanksref{Augustana}\nolinebreak,
P.~Amaudruz\thanksref{TRIUMFaddress}\nolinebreak,
M.~Atzori Corona\thanksref{CAINFN}\nolinebreak,
M.~Ave\thanksref{USP}\nolinebreak,
I.~Ch.~Avetisov\thanksref{MendeleevUniverisity}\nolinebreak
% ...
 
\begin{enumerate}[label=\textsuperscript{\arabic*}]
\item{Gran Sasso Science Institute, L'Aquila 67100, Italy\label{AQGSSI}}
\item{AstroCeNT, Nicolaus Copernicus Astronomical Center of the Polish Academy of Sciences, 00-614 Warsaw, Poland\label{AstroCeNT}}
\item{INFN Catania, Catania 95121, Italy\label{CTINFN}}
\item{Universit\`a of Catania, Catania 95124, Italy\label{CTUNI}}
\item{Instituto de F\'isica, Universidade de S\~ao Paulo, S\~ao Paulo 05508-090, Brazil\label{USP}}
\item{Pacific Northwest National Laboratory, Richland, WA 99352, USA\label{PNNLaddress}}
\item{Physics Department, Augustana University, Sioux Falls, SD 57197, USA\label{Augustana}}
\item{TRIUMF, 4004 Wesbrook Mall, Vancouver, BC V6T 2A3, Canada\label{TRIUMFaddress}}
\item{INFN Cagliari, Cagliari 09042, Italy\label{CAINFN}}
\item{Mendeleev University of Chemical Technology, Moscow 125047, Russia\label{MendeleevUniverisity}}
\item{INFN Laboratori Nazionali di Legnaro, Legnaro (Padova) 35020, Italy\label{LNLINFN}}
\item{Department of Physics, Royal Holloway University of London, Egham TW20 0EX, UK\label{RHUL}}
\item{CIEMAT, Centro de Investigaciones Energ\'eticas, Medioambientales y Tecnol\'ogicas, Madrid 28040, Spain\label{CIEMAT}}
\item{Centre de Physique des Particules de Marseille, Aix Marseille Univ, CNRS/IN2P3, CPPM, Marseille, France\label{CPPM}}
\item{INFN Napoli, Napoli 80126, Italy\label{NAINFN}}
\item{INFN Pisa, Pisa 56127, Italy\label{PIINFN}}
\item{Physics Department, Universit\`a degli Studi di Pisa, Pisa 56127, Italy\label{PIUniPHY}}
\item{INFN Sezione di Roma, Roma 00185, Italy\label{RMUnoINFN}}
\item{Physics Department, Universit\`a degli Studi di Genova, Genova 16146, Italy\label{GEUni}}
\item{INFN Genova, Genova 16146, Italy\label{GEINFN}}
\item{Physics Department, Princeton University, Princeton, NJ 08544, USA\label{Princeton}}
\item{Department of Physics, Carleton University, Ottawa, ON K1S 5B6, Canada\label{Carleton}}
\item{Physics Department, Universit\`a degli Studi di Cagliari, Cagliari 09042, Italy\label{CAUniPHY}}
\item{INFN Laboratori Nazionali del Gran Sasso, Assergi (AQ) 67100, Italy\label{AQLNGS}}
\item{Department of Electrical and Electronic Engineering, Universit\`a degli Studi di Cagliari, Cagliari 09123, Italy\label{CAUniEEE}}
\item{Physics Department, Universit\`a degli Studi ``Federico II'' di Napoli, Napoli 80126, Italy\label{NAUniPHY}}
\item{Centro de Astropart\'iculas y F\'isica de Altas Energ\'ias, Universidad de Zaragoza, Zaragoza 50009, Spain\label{Zaragoza}}
\item{Skobeltsyn Institute of Nuclear Physics, Lomonosov Moscow State University, Moscow 119234, Russia\label{MSU}}
\item{Department of Physics and Astronomy, Universit\`a degli Studi di Bologna, Bologna 40126, Italy\label{BOUniPHY}}
\item{INFN Bologna, Bologna 40126, Italy\label{BOINFN}}
\item{INFN Milano, Milano 20133, Italy\label{MIINFN}}
\item{SNOLAB, Lively, ON P3Y 1N2, Canada\label{SNOLABaddress}}
\item{Department of Physics and Astronomy, Laurentian University, Sudbury, ON P3E 2C6, Canada\label{Laurentian}}
\item{M.~Smoluchowski Institute of Physics, Jagiellonian University, 30-348 Krakow, Poland\label{Krakow}}
\item{Department of Strutture per l'Ingegneria e l'Architettura, Universit\`a degli Studi ``Federico II'' di Napoli, Napoli 80131, Italy\label{NAUniStruct}}
\item{INFN Torino, Torino 10125, Italy\label{TOINFN}}
\item{Physics Department, Sapienza Universit\`a di Roma, Roma 00185, Italy\label{RMUnoUni}}
\item{Physics Department, Universit\`a degli Studi di Salerno, Salerno 84084, Italy\label{SAUni}}
\item{INFN Salerno, Salerno 84084, Italy\label{SAINFN}}
\item{Chemistry, Physics, Mathematics and Natural Sciences Department, Universit\`a di Sassari, Sassari 07100, Italy\label{UniSS}} 
\item{Saint Petersburg Nuclear Physics Institute, Gatchina 188350, Russia\label{Petersburg}}
\item{Department of Physics, Engineering Physics and Astronomy, Queen's University, Kingston, ON K7L 3N6, Canada\label{Queens}}
\item{National Research Centre Kurchatov Institute, Moscow 123182, Russia\label{Kurchatov}}
\item{Department of Physics and Astronomy, University of California, Davis, CA 95616, USA\label{UCDavis}}
\item{Physics Department, Lancaster University, Lancaster LA1 4YB, UK\label{Lancaster}}
\item{APC, Universit\'e de Paris, CNRS, Astroparticule et Cosmologie, Paris F-75013, France\label{APC}}
\item{Museo Storico della Fisica e Centro Studi e Ricerche Enrico Fermi, Roma 00184, Italy\label{CentroFermi}}
\item{Institute for Particle Physics, ETH Z\"urich, Z\"urich 8093, Switzerland\label{ETHZ}}
\item{LPNHE, CNRS/IN2P3, Sorbonne Universit\'e, Universit\'e Paris Diderot, Paris 75252, France\label{LPNHE}}
\item{Williams College, Physics Department, Williamstown, MA 01267 USA\label{WilliamsCollege}}
\item{Department of Physics and Astronomy, University of Hawai'i, Honolulu, HI 96822, USA\label{Hawaii}}
\item{Fondazione Bruno Kessler, Povo 38123, Italy\label{TNFBK}}
\item{Trento Institute for Fundamental Physics and Applications, Povo 38123, Italy\label{TNTIFPA}}
\item{National Research Nuclear University MEPhI, Moscow 115409, Russia\label{MEPhI}}
\item{Joint Institute for Nuclear Research, Dubna 141980, Russia\label{JINR}}
\item{Institute of High Energy Physics, Beijing 100049, China\label{IHEPaddress}}
\item{Engineering and Architecture Faculty, Universit\`a di Enna Kore, Enna 94100, Italy\label{ENUniCEE}}
\item{INFN Laboratori Nazionali del Sud, Catania 95123, Italy\label{CTLNS}}
\item{Department of Physics, University of Alberta, Edmonton, AB T6G 2R3, Canada\label{Alberta}}
\item{School of Physics and Astronomy, University of Edinburgh, Edinburgh EH9 3FD, UK\label{UniversityofEdinburgh}}
\item{Universit\`a degli Studi dell’Aquila, L’Aquila 67100, Italy\label{UnivAQ}}
\item{Fermi National Accelerator Laboratory, Batavia, IL 60510, USA\label{FNALaddress}}
\item{Radiation Physics Laboratory, Belgorod National Research University, Belgorod 308007, Russia\label{Belgorod}}
\item{Pharmacy Department, Universit\`a degli Studi ``Federico II'' di Napoli, Napoli 80131, Italy\label{NAUniPHARM}}
\item{School of Physics and Astronomy, University of Birmingham, Edgbaston, B15 2TT, Birmingham, UK\label{Birmingham}}
\item{INFN Roma Tre, Roma 00146, Italy\label{RMTreINFN}}
\item{Mathematics and Physics Department, Universit\`a degli Studi Roma Tre, Roma 00146, Italy\label{RMTreUni}}
\item{Virginia Tech, Blacksburg, VA 24061, USA\label{VTech}}
\item{Fundaci\'on ARAID, Universidad de Zaragoza, Zaragoza 50009, Spain\label{ZaragozaARAID}}
\item{Physics Department, Temple University, Philadelphia, PA 19122, USA\label{Temple}}
\item{Department of Physics, University of Liverpool, The Oliver Lodge Laboratory, Liverpool L69 7ZE, UK\label{Liverpool}}
\item{University of Warwick, Department of Physics, Coventry CV47AL, UK\label{Warwick}}
\item{Budker Institute of Nuclear Physics, Novosibirsk 630090, Russia\label{BINP}}
\item{Novosibirsk State University, Novosibirsk 630090, Russia\label{NSU}}
\item{Amherst Center for Fundamental Interactions and Physics Department, University of Massachusetts, Amherst, MA 01003, USA\label{UMass}}
\item{Department of Physics and Astronomy, The University of Manchester, Manchester M13 9PL, UK\label{Manchester}}
\item{Physics Department, Universit\`a degli Studi di Milano, Milano 20133, Italy\label{MIUni}}
\item{Department of Physics, University of Houston, Houston, TX 77204, USA\label{Houston}}
\item{Chemical, Materials, and Industrial Production Engineering Department, Universit\`a degli Studi ``Federico II'' di Napoli, Napoli 80126, Italy\label{NAUniCHE}}
\item{Department of Electronics and Communications, Politecnico di Torino, Torino 10129, Italy\label{TOPoli}}
\item{Physics and Astronomy Department, University of California, Los Angeles, CA 90095, USA\label{UCLA}}
\item{University of Chinese Academy of Sciences, Beijing 100049, China\label{UCAS}}
\item{Department of Physics and Astronomy, University of California, Riverside, CA 92507, USA\label{UCRiverside}}
\item{Department of Physics and Engineering, Fort Lewis College, Durango, CO 81301, USA\label{FortLewis}}
\end{enumerate}
}

\end{document}